\documentclass[english,twocolumn,transaction]{IEEEtran}
\usepackage{float}
\usepackage{amsmath}
\usepackage{amsthm}
\usepackage{amssymb}
\usepackage{graphicx}

\makeatletter

\providecommand{\tabularnewline}{\\}
\floatstyle{ruled}
\newfloat{algorithm}{tbp}{loa}
\providecommand{\algorithmname}{Algorithm}
\floatname{algorithm}{\protect\algorithmname}

\theoremstyle{plain}
\newtheorem{thm}{\protect\theoremname}
\theoremstyle{definition}
\newtheorem{problem}{\protect\problemname}
\theoremstyle{definition}
\newtheorem{sol}{\protect\solutionname}
\theoremstyle{definition}
\newtheorem{defn}{\protect\definitionname}

\IEEEoverridecommandlockouts
\usepackage{cite}
\usepackage{amsfonts}\usepackage{algorithmic}
\usepackage{textcomp}
\usepackage{xcolor}
\usepackage{epstopdf}

\def\BibTeX{{\rm B\kern-.05em{\sc i\kern-.025em b}\kern-.08em
    T\kern-.1667em\lower.7ex\hbox{E}\kern-.125emX}}

\@ifundefined{showcaptionsetup}{}{%
 \PassOptionsToPackage{caption=false}{subfig}}
\usepackage{subfig}
\makeatother

\providecommand{\definitionname}{Definition}
\providecommand{\problemname}{Problem}
\providecommand{\solutionname}{Solution}
\providecommand{\theoremname}{Theorem}

\begin{document}

\title{Energy-Delay Minimization of Task Migration Based on Game Theory in MEC-assisted Vehicular Networks}

\author{Haipeng~Wang,~Tiejun~Lv,~\emph{Senior~Member,~IEEE},~Zhipeng~Lin,~\emph{Member,~IEEE},~and~Jie~Zeng,~\IEEEmembership{Senior Member,~IEEE}
\thanks{Manuscript received Septemper 24, 2021; revised February 20, 2022; accepted May 11, 2022. This work was supported by the National Natural Science Foundation of China (No. 62001264), and the Natural Science Foundation of Beijing (No. L192025).
\emph{(Corresponding author: Tiejun Lv and Jie Zeng).}}
\thanks{H. Wang and T. Lv are with the School of Information and Communication Engineering, Beijing University of Posts and Telecommunications (BUPT), Beijing 100876, China (e-mail: \{wanghaipeng, lvtiejun\}@bupt.edu.cn).

Z. Lin is with the Key Laboratory of Dynamic Cognitive System of Electromagnetic Spectrum Space, College of Electronic and Information Engineering, NUAA, Nanjing 211106, China (e-mail: linlzp@ieee.org).

J. Zeng is with the School of Cyberspace Science and Technology, Beijing Institute of Technology, Beijing 100081, China (e-mail: zengjie@bit.edu.cn).
}}
\maketitle
\begin{abstract}
Roadside units (RSUs), which have strong
computing capability and are close to vehicle nodes, have been widely
used to process delay- and computation-intensive tasks of
vehicle nodes. However, due to their high mobility, vehicles may drive out of the coverage of RSUs before receiving
the task processing results. In this paper, we propose
a mobile edge computing-assisted vehicular network, where vehicles can offload their tasks
to a nearby vehicle via a vehicle-to-vehicle (V2V) link or a nearby
RSU via a vehicle-to-infrastructure link. These tasks are also
migrated by a V2V link or an infrastructure-to-infrastructure (I2I)
link to avoid the scenario where the vehicles cannot receive the
processed task from the RSUs. Considering mutual interference from the same link of offloading tasks
and migrating tasks, we construct a vehicle offloading decision-based game to minimize the computation overhead. We prove that the game can always achieve Nash equilibrium and
convergence by exploiting the finite improvement property. We then propose a task migration
(TM) algorithm that includes three task-processing methods and two
task-migration methods. Based on the TM algorithm, computation overhead minimization offloading (COMO)
algorithm is presented. Extensive simulation results show that the proposed TM
and COMO algorithms reduce the computation overhead and increase the
success rate of task processing.
\end{abstract}

\begin{IEEEkeywords}
computation offloading, mobile edge computing, game theory, task migration,
I2I
\end{IEEEkeywords}

\section{Introduction}

The Internet of Vehicles (IoV), which originated from the Internet of Things (IoT), enables
vehicle-to-everything (V2X) communication to enhance intelligent driving services,
thereby improving the level of social
traffic service intelligence \cite{9,49} and the management of urban flows \cite{41}. However, due to their limited size, vehicles generally do not offer sufficient computing resources for their service requirements with low computing latency \cite{10}. As a result, efficient networks must be designed to address computing resource shortages on the vehicle side.

Considering the rich computing resources provided by the Internet, cloud-based in-car networks have been proposed to address the explosive growth of computing task requirements of vehicles.
The cloud network uses advanced communication technologies to integrate various
resources and provides assistance for task offloading. Vehicles can
either process computing tasks locally or offload tasks to the cloud.
The cloud network improves resource utilization and quality of service
(QoS) \cite{12,48}. However, with the increase in the number of smart devices, the number of computing tasks has also increased. These devices also need to offload computing tasks
to the cloud for processing, which inevitably causes network congestion
and resource waste.

Mobile edge computing (MEC) is considered the most
promising computing paradigm to improve the QoS of users. MEC effectively
integrates wireless network and Internet technologies, adds computing,
storage, processing and other functions on the wireless network side,
and uses its geographical advantages to provide more convenient services
for vehicles. MEC can improve network utilization efficiency and achieve low latency and energy consumption. In MEC networks,
vehicles can offload tasks to nearby vehicles with idle resources
or to nearby roadside units (RSUs) for processing \cite{13}. However, when
the vehicle offloads computing tasks to the RSU, due
to the high mobility of the vehicle and the increasing size of the
computing tasks, the vehicle cannot receive the processing results
of computing tasks from the RSU because the vehicle has already driven out of the RSU's communication coverage.

To receive the processing results from the
RSU, the computing tasks must be migrated from vehicles. Typically, there are two ways for vehicles to offload their tasks to an RSU with an MEC server: vehicle-to-infrastructure (V2I) mode and vehicle-to-vehicle (V2V)+V2I mode. Correspondingly, the RSU also has two
ways to migrate the computing results, i.e., infrastructure-to-infrastructure (I2I)+infrastructure-to-vehicle (I2V) mode and
I2V+V2V mode. Because the size of each offloaded computing task is different, the computing tasks require different computing and transmission
resources. Using the same channel to carry out the offloading tasks
or transmitting the computing results will cause common channel interference
\cite{14,23,24}, which will reduce the transmission/migration rate and increase energy consumption and delay. Therefore, the energy consumption and delay must be balanced, and the tasks must be completed within
a limited time.

In this paper, we propose an MEC-assisted vehicular network architecture.
We optimize the task offloading and migration
decisions in the vehicle network to minimize
the weighted sum of delay  and energy consumption. In the proposed system, we
propose a task migration (TM) algorithm and a computation overhead
minimization offloading (COMO) algorithm based on game theory. The
key contributions of this paper are summarized as follows:
\begin{itemize}
\item To reduce the computation overhead, we construct an MEC-assisted vehicular network architecture, where the co-channel interference of task offloading and task migration is considered.
\item Based on game theory, we formulate the constrained optimization
problem of offloading decisions as a game. We prove that the
game can always achieve Nash equilibrium (NE) and can converge according to the finite improvement property (FIP).
\item We propose a TM algorithm and a COMO algorithm that can reduce the computation
overhead and increase the success rate of task processing.
\end{itemize}

The rest of this paper is organized as follows: Section \mbox{II}
reviews related work, followed by the MEC-assisted vehicular network
system model in Section \mbox{III}. In Section \mbox{IV}, we construct a game-based
vehicle offloading decision and propose
a TM algorithm and a COMO algorithm. Section \mbox{V} discusses the simulation
results. Finally, we summarize our paper in Section \mbox{VI}.

\section{Related Work}

In recent years, MEC-based task offloading and task migration have
drawn the attention of researchers. The latency \cite{15,16,17,18}
and the energy consumption of the system \cite{19,20,21,22} are
two criteria used to evaluate the performance of task offloading and
task migration.

\subsection{Cloud-Based Computation Offloading }

Since task offloading can effectively alleviate the burden of
insufficient vehicle computing resources, cloud-based computation
offloading has been applied to various situations. In vehicular fog
and cloud computing (VFCC) systems, to address the impact of vehicles
leaving unfinished tasks, the authors in \cite{27} described the task
offloading problem as a semi-Markov decision process (SMDP), and the
value iteration algorithm of the SMDP was designed to maximize the total
long-term benefits of the VFCC system. To effectively improve
the offloading efficiency, a method of combining resource allocation
and offloading was proposed in \cite{28}, where a low-complexity algorithm
was developed to jointly optimize the offloading decision and resource
allocation. Wang et al. \cite{29} formulated the offloading problem
as an optimization problem and proposed a fog-cloud computational offloading algorithm and a heuristic algorithm to jointly minimize the
power consumption of vehicles and that of the computational facilities.

\subsection{Fog-Based Computation Offloading}

Fog computing, as a transition between cloud computing and edge
computing, concentrates data, processing, and applications in devices at the edge of the network, rather than storing them in the cloud. To detect and take necessary steps
for public safety during a disaster, the authors in \cite{32} proposed
crowdsourcing-based disaster management using a fog computing
model in the IoT. Zhang et al. \cite{31} proposed a regional cooperative
fog-computing-based intelligent vehicular network architecture
and a hierarchical model with intra-fog and inter-fog resource management
to optimize the energy efficiency and packet dropping rates. To minimize the average offloading delay, the task offloading problem of dynamic fog networks was strictly defined as an online stochastic optimization problem. In \cite{30}, the authors designed task offloading strategies for stationary status and nonstationary status algorithms and proposed the use of a discount factor to reduce the delay.

\subsection{MEC-Based Computation Offloading }

In the MEC system, vehicles can offload computation tasks to other
devices that have strong computing power and resources to reduce
the system energy consumption and delay \cite{1}. Considering the
mutual interference of tasks in the same channel, the authors in \cite{2}
and \cite{3} constructed MEC-assisted networks and proposed algorithms
based on game theory to minimize the network computation overhead.
To meet the requirements of mobility and energy harvesting of an IoT network,
the authors in \cite{4} proposed an online mobility-aware offloading
and a resource allocation algorithm that can balance the system service
cost and energy queue length. An efficient algorithm based on
submodular optimization and a collaborative task computing scheme
were proposed in \cite{5} to efficiently
mitigate data redundancy, conserve network bandwidth consumption and
reduce the cost of processing tasks.

\subsection{Task Migration}

With the development of science and technology, the speed of vehicles has substantially improved, and the size of the data that needs to be
processed has increased. To effectively address the resulting tasks from vehicle offloading, efficient task migration strategies must be designed such that vehicles can
offload their tasks to the RSU for processing, where the
RSU determines the task processing mode by combining the delay threshold and energy consumption requirements
of the task.

\subsubsection{Task migration in Traditional Optimization Mode}
In \cite{7}, to achieve an optimal
balance between energy consumption and time cost during service migration,
the authors designed a 6th generation mobile network (6G)-enabled box task migration
method and a strength Pareto evolutionary algorithm for the IoV. The authors in \cite{40} establish an efficient service migration
model and build a nonlinear 0-1 programming problem by designing
a business migration scheme based on a particle swarm, effectively
reducing the latency and energy consumption. In \cite{8}, by introducing distributed traffic guidance in large MEC systems and distinguishing between two different types of network elements, the scalability problem of a large MEC network was resolved as a partitioned MEC network, namely, edge servers and routers. Then, dynamic shortest path selection and dynamic multipath searching algorithms were proposed to achieve high QoS with ultra-low latency.

\subsubsection{Task Migration in Deep Learning Mode}

In \cite{39}, the authors propose a multi-agent
deep reinforcement learning algorithm that maximizes the comprehensive
utility of communication, computing and routing planning in a distributed
manner, thus reducing service delay, migration cost and travel time. The authors in \cite{6} proposed a deep Q-learning service migration decision
algorithm and a neural network-based service migration framework that realizes the adaptive migration of task offloading when connected vehicles move. To minimize the data processing cost within the system and ensure the delay constraints of applications, the author in \cite{38}
formulates a unified communication, computing, caching and collaborative
computing framework and develops a cooperative data scheduling scheme
to model the data scheduling as a deep reinforcement learning problem
which is solved by an enhanced deep Q-Network algorithm with
a single target Q-network.

Different from these aforementioned studies, our work considers the mutual interference of tasks in the same channel, whenever the channel processes the common channel offloading tasks or the common channel transmitting computing results. On this basis, a TM algorithm
and a COMO algorithm based on game theory are proposed to reduce the computation overhead and increase the success rate of task processing according to the offloading decisions of the vehicle.

\section{System Model and Problem Formulation }

In this section, we first introduce the MEC-assisted vehicular network.
Then, we analyze the weighted sum of delay and energy consumption when the vehicles are offloading and migrating tasks.

\subsection{System Model}

\begin{figure}
\centering

\includegraphics[width=8cm]{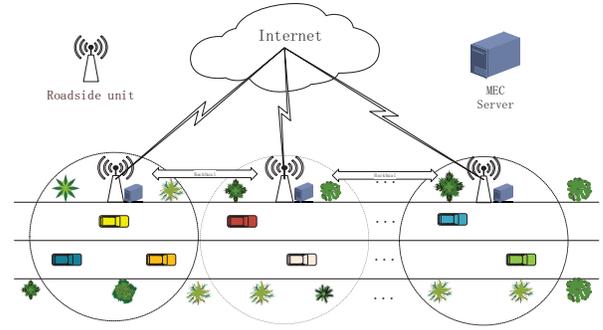}

\caption{An MEC-assisted vehicular network system.\label{fig:A-MEC-assisted-V2X}}
\end{figure}

As shown in Fig. \ref{fig:A-MEC-assisted-V2X}, we consider an MEC-assisted
vehicular network system with multiple vehicles and $S$ RSUs, where the
set of RSUs is $\mathbf{RSU}=\{RSU_{1},RSU_{2},...,RSU_{S}\}$.
In the proposed network, we assume that vehicles are
equipped with 802.11p interfaces and network interfaces. Every
vehicle is equipped with on-board units (OBUs) \cite{11}, GPS, wireless communication devices and other devices. GPS can provide real-time information about the current positions and directions of vehicles \cite{42}. The OBU is a microwave device that uses dedicated short-range communication technology to communicate with the RSU: the OBU has limited computing power and storage capacity to process computing tasks with small data size. The RSUs, which have strong computing power, are used as edge servers.

In the MEC-assisted vehicular network, we use software
defined networks (SDN) technology. An SDN, which has deep programmability, has a physical data plane and an abstracted control plane \cite{46}. In the network, the sub-SDN controller collects idle vehicle information and target vehicle computing task information. Idle vehicle information includes vehicle location, driving direction, and CPU frequency of the vehicle server. The target vehicle calculation task information includes task size, calculation workload and maximum delay tolerance. The vehicle usually periodically sends the information to the nearest RSU. The RSUs store the collected information in the data center for invocation by the SDN controller \cite{45}.

According to the unit circle protocol model \cite{25}, the communication
range of each RSU is a circle with radius $\frac{L_{\mathrm{I2I}}}{2}$.
In the system, we use a single-input and single-output orthogonal
frequency division multiple access scheme to avoid unnecessary
interference between RSUs. The set of vehicles is $\mathbf{M}=\{1,2,...,M\}$,
and we use $m$ to denote a vehicle that has a computation-intensive
or delay-sensitive task.

In the system, the V2V communication link, the V2I communication link,
the V2V migration link and the I2I migration link are deployed at
different frequencies; thus, they do not interfere with each other. There
are three ways for vehicles to process their tasks: 1) process
tasks locally; 2) offload tasks to idle vehicles nearby via
a V2V communication link; or 3) offload tasks to the nearby RSUs for processing via a V2I communication link. Moreover, there are two
ways for RSUs to migrate the computing results: 1) RSU $i$ uses the I2I migration link to transmit the computing results to RSU $q$ where vehicle $m$ resides and then RSU $q$ transmits
the computing results for vehicle $m$ through the I2V communication
link; or 2) RSU $i$ uses the I2V communication link to transmit
the computing results to vehicle $e$ in its communication range
and vehicle $e$ transmits the computing results for vehicle
$m$ through the V2V migration link.

Note that the task of vehicle $m$ cannot be split. In the system, we use the
set $\left\{ D_{m,\mathrm{in}},C_{m,\mathrm{in}},\tau\right\} $ to
denote the task of vehicle $m$. $D_{m,\mathrm{in}}$ (in bits) is
the size of the data to be processed by the task, $C_{m,\mathrm{in}}$
(in CPU cycles) is the number of cycles required to process the task by
the CPU, and $\tau$ (in s) is the maximum delay tolerance for the
task. $C_{m,\mathrm{in}}=\delta D_{m,\mathrm{in}}$, where $\delta$ is the number of CPU cycles required for processing one bit of data. Vectors of $C_{m,\mathrm{in}}$ and $D_{m,\mathrm{in}}$ are denoted as $\mathbf{D_\mathbf{in}}=\left\{ D_{1,\mathrm{in}},D_{2,\mathrm{in}},...,D_{M,\mathrm{in}}\right\} $
and $\mathbf{C_\mathbf{in}}=\left\{ C_{1,\mathrm{in}},C_{2,\mathrm{in}},...,C_{M,\mathrm{in}}\right\} $. In
this paper, the energy consumption of the vehicles and the delay of
the system are considered. In some jobs, such as data redundancy processing, the amount of returned data is still very large, and thus cannot be ignored. Note that due to the large size of the feedback
data $D_{m,\mathrm{out}}$, the energy consumption and delay of downlink transmission must be considered. $D_{m,\mathrm{in}}=\xi D_{m,\mathrm{out}}$, where $\xi$ is the reciprocal of the data processing coefficient of the RSU.
For vehicle $m$, there are three ways to handle tasks: local processing,
V2V offloading and V2I offloading. Furthermore, the system has two ways to
migrate tasks, i.e., V2V migration and I2I migration. We use $\mathbf{\Psi}=\{d_{m}=0,1,2,3,4\}$
to denote the offloading decision of vehicle $m$. Specifically, when
$d_{m}=0$, vehicle $m$ processes the task locally; when $d_{m}=1$,
vehicle $m$ offloads the task to an idle vehicle nearby for processing;
when $d_{m}=2$, vehicle $m$ offloads the task to an RSU for processing
and transfers the results directly; when $d_{m}=3$, vehicle
$m$ offloads the task to an RSU for processing and uses V2V migration
to transfer the computing results; and when $d_{m}=4$, vehicle
$m$ offloads the task to an RSU for processing and uses I2I migration
to transfer the computing results. Co-channel interference occurs when vehicles use the same subcarrier offloading tasks.

\subsubsection{Local Processing}

In the considered system, we assume that the vehicles are distributed according to a Poisson distribution
on the highway and that the vehicles move at an average speed of $\bar{\upsilon}$. Let $\sigma$ be the probability density of finding a vehicle
per meter. Then, the probability of finding $n$ vehicles in
the $l_{\mathrm{V2V}}$-meter lane can be expressed as
\begin{equation}
f(n,l_{\mathrm{V2V}})=\frac{(\sigma l_{\mathrm{V2V}})^{n}}{n!}e^{-\sigma l_{\mathrm{V2V}}},\:n\geq0.
\end{equation}

We use $L_{\mathrm{V2V}}$ to express the distance between two vehicles. The probability that the distance between the vehicles is less than $l_{\mathrm{V2V}}$, is written as
\begin{equation}
\mathrm{Pr}\{L_{\mathrm{V2V}}\leq l_{\mathrm{V2V}}\}=1-e^{-\sigma l_{\mathrm{V2V}}}.\label{eq:0}
\end{equation}

According to \eqref{eq:0}, $L_{\mathrm{V2V}}$ is
independently identically distributed and conforms to an exponential
distribution. Moreover, we can obtain the expected value $\frac{1}{\sigma}$ of
the distance between the vehicles, which is treated as the average
distance $L_{\mathrm{V2V}}$ between two vehicles, i.e., $L_{\mathrm{V2V}}=\frac{1}{\sigma}$.

In the system, we modeled the CPU energy consumption of the vehicle
as $P_{exe}=k\mu_{m}^{3}f_{\mathrm{ue}}^{3}$ as in \cite{34}, where
$\mathrm{\mathbf{\mathit{f}}_{ue}}$, $k$ and $\mu_{m}\in(0,1]$
denote the CPU cycle frequency of vehicle $m$, the energy coefficient
depending on the chip architecture \cite{33} and the available rate
of computational resources, respectively.

When vehicle $m$ processes its tasks locally, the delay $T_{m}^{\mathrm{local}}$
of vehicle $m$ is given by
\begin{equation}
T_{m}^{\mathrm{local}}=\frac{C_{m,\mathrm{in}}}{\mu_{m}\mathrm{\mathit{f}}_{\mathrm{ue}}},\label{eq:local_delay}
\end{equation}
where $\mu_{m}\in(0,1]$ is the available rate of computational resources.
Correspondingly, the execution consumption $E_{m}^{\mathrm{local}}$
of vehicle $m$ is expressed as
\begin{equation}
E_{m}^{\mathrm{local}}=P_{exe}T_{m}^{\mathrm{local}}=k\mu_{m}^{2}C_{m,\mathrm{in}}\mathrm{\mathit{f}_{ue}^{2}}.\label{eq:local_energy}
\end{equation}

When we consider the weighted sum of delay and energy consumption,
the local computation overhead $\Omega_{m}^{\mathrm{local}}$ can
be obtained as
\begin{equation}
\Omega_{m}^{\mathrm{local}}=\alpha_{m}T_{m}^{\mathrm{local}}+\beta_{m}E_{m}^{\mathrm{local}},\label{eq:local_computation_overhead}
\end{equation}
where $\alpha_{m}$ and $\beta_{m}$ denote the weights of delay and
energy consumption of vehicle $m$, respectively. $\alpha_{m},\beta_{m}\in[0,1]$,
$\alpha_{m}+\beta_{m}=1$. If $\alpha_{m}>\beta_{m}$, vehicle $m$ needs to process a delay-sensitive task; and
if $\beta_{m}>\alpha_{m}$, vehicle $m$ needs
to process a computational-intensive task. When $\alpha_{m}=1$ or
$\beta_{m}=1$, vehicle $m$ is concerned with only
the delay or energy consumption.

\subsubsection{V2V Processing}

V2V processing includes three main operations: task
offloading, task processing and computation result
returning. The vehicle offloads the task to a nearby idle vehicle; then, the idle
vehicle processes the computing task and delivers the computing
results to the target vehicle. If the vehicle decides to offload the
task to an idle vehicle, this vehicle can offload the task via the
V2V communication link. This process can lead to task-transmission delays and energy consumption. We use $R_{\mathrm{\mathit{m},tran}}^{\mathrm{V2V}}(\boldsymbol{d})\text{ }$
to denote the transmission rate of a V2V communication link. Since the same channel
is used to transmit the computing result and the offloading task. It is reasonable to assume that their transmit rates are equal. Let $p_{\mathrm{\mathit{m},tran}}^{\mathrm{V2V}}$
and $\mathrm{b_{\mathit{m},tran}^{V2V}}$ denote the transmission
power and the bandwidth of the V2V communication
link, respectively. According to the offloading decisions $\boldsymbol{d}=\{d_{1},d_{2},d_{3},...,d_{M}\}$,
the transmission rate of vehicle $m$ can be obtained. The transmission
rate can be expressed as
\begin{equation}
R_{\mathrm{\mathit{m},tran}}^{\mathrm{V2V}}(\boldsymbol{d})=\mathrm{\mathrm{b_{\mathit{m},tran}^{V2V}}\log_{2}(1+\gamma_{\mathit{m},tran}^{V2V}),}
\end{equation}
where $\mathrm{\gamma_{\mathit{m},tran}^{V2V}}$ is the signal-to-interference-plus-noise ratio (SINR) of the V2V communication link. If
there is no V2V communication link sharing, we have $\mathrm{\gamma_{\mathit{m},tran}^{V2V}}=p_{\mathrm{\mathit{m},tran}}^{\mathrm{V2V}}h_{\mathrm{\mathit{m},tran}}^{\mathrm{V2V}}/\omega$,
where $\omega$ denotes the power of complex Gaussian white
noise and $h_{\mathrm{\mathit{m},tran}}^{\mathrm{V2V}}$ denotes the
channel gain between adjacent vehicles that are offloading tasks through a V2V communication link. If
V2V communication link sharing occurs, the SINR of the V2V communication link
can be expressed as
\begin{equation}
\mathrm{\gamma_{\mathit{m},tran}^{V2V}}=\frac{p_{\mathrm{\mathit{m},tran}}^{\mathrm{V2V}}h_{\mathrm{\mathit{m},tran}}^{\mathrm{V2V}}}{\omega+\sum_{g\in\mathbf{M},g\neq m,d_{g}=1}^{M}p_{\mathrm{\mathit{g},tran}}^{\mathrm{V2V}}h_{\mathrm{\mathit{g},tran}}^{\mathrm{V2V}}}.
\end{equation}

When a vehicle offloads its task to a nearby idle vehicle, we can
obtain the delay $T_{m}^{\mathrm{V2V}}(\boldsymbol{d})$, given
by
\begin{equation}
T_{m}^{\mathrm{V2V}}(\boldsymbol{d})=\frac{D_{m,\mathrm{in}}}{R_{\mathrm{\mathit{m},tran}}^{\mathrm{V2V}}(\boldsymbol{d})}+\frac{C_{m,\mathrm{in}}}{\mathit{f}_{\mathrm{ue}}}+\frac{D_{m,\mathrm{out}}}{R_{\mathrm{\mathit{m},tran}}^{\mathrm{V2V}}(\boldsymbol{d})}.\label{eq:V2V_delay}
\end{equation}

Similarly, we can obtain the energy consumption $E_{m}^{\mathrm{V2V}}(\boldsymbol{d})$,
given by
\begin{align}
 & E_{m}^{\mathrm{V2V}}(\boldsymbol{d})\nonumber \\
 & =\frac{p_{m,\mathrm{tran}}^{\mathrm{V2V}}D_{m,\mathrm{in}}}{R_{m,\mathrm{tran}}^{\mathrm{V2V}}(\boldsymbol{d})}+kC_{m,\mathrm{in}}\mathit{f}_{\mathrm{ue}}^{2}+\frac{p_{m,\mathrm{tran}}^{\mathrm{V2V}}D_{m,\mathrm{out}}}{R_{m,\mathrm{tran}}^{\mathrm{V2V}}(\boldsymbol{d})}\text{.}\label{eq:V2V_energy}
\end{align}

According to \eqref{eq:local_computation_overhead}, we can obtain computation overhead $\Omega_{m}^{\mathrm{V2V}}(\boldsymbol{d})$
of the two vehicles, given by
\begin{equation}
\Omega_{m}^{\mathrm{V2V}}(\boldsymbol{d})=\alpha_{m}T_{m}^{\mathrm{V2V}}(\boldsymbol{d})+\beta_{m}E_{m}^{\mathrm{V2V}}(\boldsymbol{d}).\label{eq:V2V_computation_overhead}
\end{equation}

According to \eqref{eq:V2V_computation_overhead}, when the number of vehicles using V2V communication links increases,
the transmission rate of vehicles decrease; accordingly,
the transmission time and energy consumption increase. If vehicle
$m$ needs to process a delay-sensitive task, a larger $\alpha_{m}$
can be taken. If vehicle $m$ needs to process a computation-intensive
task, a larger $\beta_{m}$ can be taken.

\subsubsection{V2I Processing}

When a vehicle decides to offload its task to a nearby RSU, it can
use the V2I communication link. The RSU uses its powerful computing
power to process the computing task and then transmit the results back
to the target vehicle. Considering that the computing result cannot
be ignored, we have to calculate the resulting transmission delay
and transmission energy consumption. In the system, we use $R_{m,\mathrm{tran}}^{\mathrm{V2I}}(\boldsymbol{d})$
to denote the transmission rate between vehicle $m$ and the nearby RSU that establishes the V2I communication link. Considering the
same channel is used to transmit the computing result and the offloading
task, their transmit rates are equal. According to the offloading
decisions $\boldsymbol{d}=(d_{1},d_{2},d_{3},...,d_{M})$, the transmission
rate of the V2I communication link can be obtained. The transmission rate
can be expressed as
\begin{equation}
R_{m,\mathrm{tran}}^{\mathrm{V2I}}(\boldsymbol{d})=\mathrm{b_{\mathit{m},tran}^{V2I}}\log_{2}(1+\mathrm{\gamma_{\mathit{m},tran}^{V2I}}),
\end{equation}
where $\mathrm{b_{\mathit{m},tran}^{V2I}}$ denotes the bandwidth
of the V2I communication link; and $\mathrm{\gamma_{\mathit{m},tran}^{V2I}}$
is the SINR of the V2I communication link. If there is no V2I communication
link sharing, $\mathrm{\gamma_{\mathit{m},tran}^{V2I}}=p_{\mathrm{\mathit{m},tran}}^{\mathrm{V2I}}h_{\mathrm{\mathit{m},tran}}^{\mathrm{V2I}}/\omega$,
where $p_{m,\mathrm{tran}}^{\mathrm{V2I}}$ denotes the transmission
power of the V2I communication link and $h_{m,\mathrm{tran}}^{\mathrm{V2I}}$
denotes the channel gain between vehicle $m$ and the nearby RSU.
If there is V2V communication link sharing, the transmission rate is affected by other vehicles that choose the same V2I communication link to offload their tasks. As a result, interference will exist during the data transmission phase. The SINR of the V2V communication link can be expressed as
\begin{equation}
\mathrm{\gamma_{\mathit{m},tran}^{V2I}}=\frac{p_{m,\mathrm{tran}}^{\mathrm{V2I}}h_{m,\mathrm{tran}}^{\mathrm{V2I}}}{\omega+\sum_{g\in\mathbf{M},g\neq m,d_{g}=2,3,4}^{M}p_{g,\mathrm{tran}}^{\mathrm{V2I}}h_{g,\mathrm{tran}}^{\mathrm{V2I}}}.
\end{equation}

In the task migration model, computing tasks are all processed on the RSU, so the interference increases when $d_{m}=2$, $d_{m}=3$,
and $d_{m}=4$ are applied.

When a vehicle offloads its task to an idle RSU, we can obtain
the delay $T_{m}^{\mathrm{V2I}}(\boldsymbol{d})$, given by
\begin{equation}
T_{m}^{\mathrm{V2I}}(\boldsymbol{d})=\frac{D_{m,\mathrm{in}}}{R_{m,\mathrm{tran}}^{\mathrm{V2I}}(\boldsymbol{d})}+\frac{C_{m,\mathrm{in}}}{\mathrm{\mathit{f}_{mec}}}+\frac{D_{m,\mathrm{out}}}{R_{m,\mathrm{tran}}^{\mathrm{V2I}}(\boldsymbol{d})},\label{eq:V2I_delay}
\end{equation}
where $\mathrm{\mathbf{\mathit{f}}_{mec}}$ denotes the CPU cycle
frequency of the RSU.

Similarly, we can obtain the energy consumption $E_{m}^{\mathrm{V2I}}(\boldsymbol{d})$,
given by
\begin{equation}
E_{m}^{\mathrm{V2I}}(\boldsymbol{d})=p_{m,\mathrm{tran}}^{\mathrm{V2I}}\frac{D_{m,\mathrm{in}}}{R_{m,\mathrm{tran}}^{\mathrm{V2I}}(\boldsymbol{d})}+p_{m,\mathrm{tran}}^{\mathrm{V2I}}\frac{D_{m,\mathrm{out}}}{R_{m,\mathrm{tran}}^{\mathrm{V2I}}(\boldsymbol{d})}.\label{eq:V2I_energy}
\end{equation}

According to \eqref{eq:local_computation_overhead}, we can obtain computation overhead $\Omega_{m}^{\mathrm{V2I}}(\boldsymbol{d})$
of vehicle $m$ and $\mathrm{RSU}$, given by
\begin{equation}
\Omega_{m}^{\mathrm{V2I}}(\boldsymbol{d})=\alpha_{m}T_{m}^{\mathrm{V2I}}(\boldsymbol{d})+\beta_{m}E_{m}^{\mathrm{V2I}}(\boldsymbol{d}).\label{eq:V2I_computation_overhead}
\end{equation}

\subsubsection{V2V Migration}

In this section, we discuss the transmission of
computing results through V2V migration links. Considering some offloading tasks with a large amount of data or high
time requirements, coupled with the fast speed of vehicles on the expressway,
vehicles will often drive out of the RSU communication range before
receiving the computing results. In this case, task migration technology
is needed to transmit the computing results. The system adopts orthogonal
frequency division multiple access for channel access. Thus,
there is no interference between the V2V communication link and the
V2V migration link. In the
system, $R_{m,\mathrm{mig}}^{\mathrm{V2V}}(\boldsymbol{d}\text{)}$
denotes the migration rate between the two vehicles. According to
the offloading decisions $\boldsymbol{d}=\{d_{1},d_{2},d_{3},...,d_{M}\}$,
we can obtain the V2V migration rate. The migration rate can be expressed
as
\begin{equation}
R_{m,\mathit{mig}}^{\mathrm{V2V}}(\boldsymbol{d})=\mathrm{b_{\mathit{m},mig}^{V2V}}\log_{2}(1+\mathrm{\gamma_{\mathit{m},mig}^{V2V}}),
\end{equation}
where $\mathrm{b_{\mathit{m},mig}^{V2V}}$ and $\mathrm{\gamma_{\mathit{m},tran}^{V2V}}$
denote the bandwidth of the V2V migration link and the SINR of the V2V migration
link, respectively. If there is no V2V migration link sharing, $\mathrm{\gamma_{\mathit{m},mig}^{V2V}}=p_{\mathrm{\mathit{m},mig}}^{\mathrm{V2V}}h_{\mathrm{\mathit{m},mig}}^{\mathrm{V2V}}/\omega$,
where $p_{m,\mathrm{mig}}^{\mathrm{V2V}}$ denotes the transmission
power of the V2V migration link and $h_{\mathrm{\mathit{m},mig}}^{\mathrm{V2V}}$
denotes the channel gain between adjacent vehicles that have
established the V2V migration link when transmitting computing results.
If there is V2V migration link sharing, it will be interfered with by other
vehicles that choose the same V2V migration link to transmit the
computing results. As a result, interference will exist during the
data transmission phase. The SINR of the V2V migration link can be expressed as
\begin{equation}
\mathrm{\gamma_{\mathit{m},mig}^{V2V}}=\frac{p_{m,\mathrm{mig}}^{\mathrm{V2V}}h_{m,\mathrm{mig}}^{\mathrm{V2V}}}{\omega+\sum_{g\in\mathbf{M},g\neq m,d_{g}=3}^{M}p_{g,\mathrm{mig}}^{\mathrm{V2V}}h_{g,\mathrm{mig}}^{\mathrm{V2V}}}.
\end{equation}

\begin{figure}
\centering\includegraphics[width=8cm]{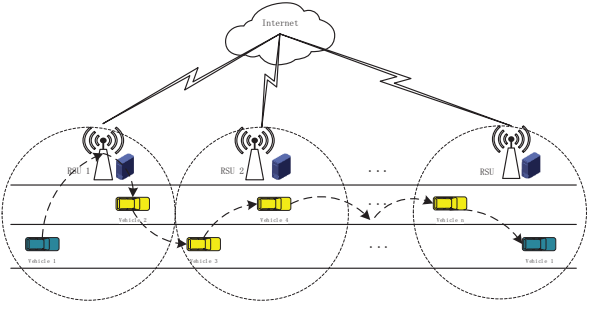}\caption{The task offloading via V2V migration link.\label{fig:The-offloading-via} }
\end{figure}

For V2V offloading, we can obtain the delay $T_{m,\mathrm{mig}}^{\mathrm{V2V,}}$
and the energy consumption $E_{m,\mathrm{mig}}^{\mathrm{V2V}}(\boldsymbol{d})$,
given by
\begin{align}
 & T_{m,\mathrm{mig}}^{\mathrm{V2V}}(\boldsymbol{d})\nonumber \\
 & =T_{m}^{\mathrm{V2I}}(\boldsymbol{d})+\phi\frac{D_{m,\mathrm{out}}}{R_{m,\mathrm{mig}}^{\mathrm{V2V}}(\boldsymbol{d})}\nonumber \\
 & =\frac{D_{m,\mathrm{out}}+D_{m,\mathrm{in}}}{R_{m,\mathrm{tran}}^{\mathrm{V2I}}(\boldsymbol{d})}+\frac{C_{m,\mathrm{in}}}{\mathit{f}_{\mathrm{mec}}}+\phi\frac{D_{m,\mathrm{out}}}{R_{m,\mathrm{mig}}^{\mathrm{V2V}}(\boldsymbol{d})},\label{eq:V2V_migration_delay}
\end{align}
\begin{align}
 & E_{m,\mathrm{mig}}^{\mathrm{V2V}}(\boldsymbol{d})\nonumber \\
 & =\frac{p_{m,\mathrm{tran}}^{\mathrm{V2I}}D_{m,\mathrm{in}}}{R_{m,\mathrm{tran}}^{\mathrm{V2I}}(\boldsymbol{d})}+\frac{p_{m,\mathrm{tran}}^{\mathrm{V2I}}D_{m,\mathrm{out}}}{R_{m,\mathrm{tran}}^{\mathrm{V2I}}(\boldsymbol{d})}+\phi\frac{p_{m,\mathrm{mig}}^{\mathrm{V2V}}D_{m,\mathrm{out}}}{R_{m,\mathrm{mig}}^{\mathrm{V2V}}(\boldsymbol{d})},\label{eq:V2V_migration_energy}
\end{align}
where $\phi$ denotes the number of V2V migration vehicles that passed
when the target vehicle received the computing result. Fig. \ref{fig:The-offloading-via} clearly illustrates the process
of the V2V task migration. When a vehicle processes a task in V2V migration
mode, the target vehicle does not receive the computing
result from the RSU within the limited time. Therefore, the vehicle
offloads the task to a nearby RSU through the V2I communication
link. The RSU processes the task and then transmits the computing
result back to a vehicle that is not the target vehicle in its coverage
area. After migration, the vehicle transmits the computing result
to the target vehicle.

\begin{problem}
When the target vehicle received the computing result, how to compute the value of $\phi$ that the number of V2V migrations that the target vehicle passed?
\end{problem}
\begin{sol}
As shown in Fig. \ref{fig:The-offloading-via}, we know that the distance of the target vehicle driven is greater than $\phi$ times the average distance between vehicles and less than $\phi$ times the average distance between vehicles plus the coverage diameter of RSU within the time when the target vehicle obtains the computing results. Thus, we can obtain
\[
\phi L_{\mathrm{V2V}}<\bar{\upsilon}T_{m,\mathrm{mig}}^{\mathrm{V2V}}(\boldsymbol{d})<\phi L_{\mathrm{V2V}}+L_{\mathrm{I2I}},
\]
\[
\Rightarrow\begin{cases}
\phi L_{\mathrm{V2V}}<\bar{\upsilon}(T_{m}^{\mathrm{V2I}}(\boldsymbol{d})+\phi\frac{D_{m,\mathrm{out}}}{R_{m,\mathrm{mig}}^{\mathrm{V2V}}(\boldsymbol{d})}),\\
\bar{\upsilon}(T_{m}^{\mathrm{V2I}}(\boldsymbol{d})+\phi\frac{D_{m,\mathrm{out}}}{R_{m,\mathrm{mig}}^{\mathrm{V2V}}(\boldsymbol{d})})<\phi L_{\mathrm{V2V}}+L_{\mathrm{I2I}},
\end{cases}
\]
\[
\Rightarrow\begin{cases}
\phi>\frac{\bar{\upsilon}(\frac{D_{m,\mathrm{in}}}{R_{m,\mathrm{tran}}^{\mathrm{V2I}}(\boldsymbol{d})}+\frac{C_{m,\mathrm{in}}}{\mathit{f}_{\mathrm{mec}}}+\frac{D_{m,\mathrm{out}}}{R_{m,\mathrm{tran}}^{\mathrm{V2I}}(\boldsymbol{d})})-L_{\mathrm{I2I}}}{L_{\mathrm{V2V}}-\bar{\upsilon}\frac{D_{m,\mathrm{out}}}{R_{m,\mathrm{mig}}^{\mathrm{V2V}}(\boldsymbol{d})}},\\
\phi<\frac{\bar{\upsilon}(\frac{D_{m,\mathrm{in}}}{R_{m,\mathrm{tran}}^{\mathrm{V2I}}(\boldsymbol{d})}+\frac{C_{m,\mathrm{in}}}{\mathit{f}_{\mathrm{mec}}}+\frac{D_{m,\mathrm{out}}}{R_{m,\mathrm{tran}}^{\mathrm{V2I}}(\boldsymbol{d})})}{L_{\mathrm{V2V}}-\bar{\upsilon}\frac{D_{m,\mathrm{out}}}{R_{m,\mathrm{mig}}^{\mathrm{V2V}}(\boldsymbol{d})}}\text{,}
\end{cases}
\]
subject to
\begin{align*}
 & L_{\mathrm{I2I}}<\bar{\upsilon}(\frac{D_{m,\mathrm{in}}}{R_{m,\mathrm{tran}}^{\mathrm{V2I}}(\boldsymbol{d})}+\frac{C_{m,\mathrm{in}}}{\mathit{f}_{\mathrm{mec}}}+\frac{D_{m,\mathrm{out}}}{R_{m,\mathrm{tran}}^{\mathrm{V2I}}(\boldsymbol{d})}),\\
 & L_{\mathrm{V2V}}>\bar{\upsilon}\frac{D_{m,\mathrm{out}}}{R_{m,\mathrm{mig}}^{\mathrm{V2V}}(\boldsymbol{d})}.
\end{align*}

In the system, we express the average of $\bar{\phi}$ as
\[
\bar{\phi}=\lfloor\frac{2\bar{\upsilon}(\frac{D_{m,\mathrm{in}}}{R_{m,\mathrm{tran}}^{\mathrm{V2I}}(\boldsymbol{d})}+\frac{C_{m,\mathrm{in}}}{\mathit{f}_{\mathrm{mec}}}+\frac{D_{m,\mathrm{out}}}{R_{m,\mathrm{tran}}^{\mathrm{V2I}}(\boldsymbol{d})})-L_{\mathrm{I2I}}}{2(L_{\mathrm{V2V}}-\bar{\upsilon}\frac{D_{m,\mathrm{out}}}{R_{m,\mathrm{mig}}^{\mathrm{V2V}}(\boldsymbol{d})})}\rfloor,
\]where $\lfloor.\rfloor$ denotes rounding down.

According to \eqref{eq:local_computation_overhead}, we can obtain computation overhead $\Omega_{m,\mathrm{mig}}^{\mathrm{V2V}}(\boldsymbol{d})$
of migrating computing results via the V2V migration link, as
\begin{equation}
\Omega_{m,\mathrm{mig}}^{\mathrm{V2V}}(\boldsymbol{d})=\alpha_{m}T_{m,\mathrm{mig}}^{\mathrm{V2V}}(\boldsymbol{d})+\beta_{m}E_{m,\mathrm{mig}}^{\mathrm{V2V}}(\boldsymbol{d}).\label{eq:V2V_migration_computation_overhead}
\end{equation}
\end{sol}

\subsubsection{I2I Migration}

In this system, we consider the communication link between adjacent
RSUs. We use this link as a task migration link to transmit the computing
results. In this section, we discuss the transmission of computing
results via I2I migration links. Let $R_{m,\mathrm{mig}}^{\mathrm{I2I}}(\boldsymbol{d}\text{)}$ denote
the migration rate between two RSUs. According to the offloading
decisions $\boldsymbol{d}=\{d_{1},d_{2},d_{3},...,d_{M}\}$, we can
obtain the I2I migration rate. The migration rate can be expressed
as
\begin{equation}
R_{\mathrm{\mathit{m},mig}}^{\mathrm{I2I}}(\boldsymbol{d})=\mathrm{b_{\mathit{m},mig}^{I2I}}\log_{2}(1+\mathrm{\gamma_{\mathit{m},mig}^{I2I}}),
\end{equation}
where $\mathrm{b_{\mathit{m},mig}^{I2I}}$ and $\mathrm{\gamma_{\mathit{m},tran}^{I2I}}$
denote the bandwidth and SINR of I2I migration
link, respectively. If there is no I2I migration link sharing, $\mathrm{\gamma_{\mathit{m},mig}^{I2I}}=p_{\mathrm{\mathit{m},mig}}^{\mathrm{I2I}}h_{\mathrm{\mathit{m},mig}}^{\mathrm{I2I}}/\omega$,
where $p_{m,\mathrm{mig}}^{\mathrm{I2I}}$ denotes the transmission
power of the I2I migration link and $h_{\mathrm{\mathit{m},mig}}^{\mathrm{I2I}}$
denotes the channel gain between adjacent RSUs that have established
the I2I migration link when transmitting computing results. If there is
I2I migration link sharing, the transmission rate will be interfered with by other vehicles
that choose the same I2I migration link to transmit the computing
results. As a result, interference will exist during the data transmission
phase. The SINR of the I2I migration link can be expressed as
\begin{equation}
\mathrm{\gamma_{\mathit{m},mig}^{I2I}}=\frac{p_{m,\mathrm{mig}}^{\mathrm{I2I}}h_{m,\mathrm{mig}}^{\mathrm{I2I}}}{\omega+\sum_{g\in\mathbf{M},g\neq m,d_{g}=4}^{M}p_{\mathrm{\mathit{g},mig}}^{\mathrm{I2I}}h_{\mathrm{\mathit{g},mig}}^{\mathrm{I2I}}}.
\end{equation}

\begin{figure}
\centering\includegraphics[width=8cm]{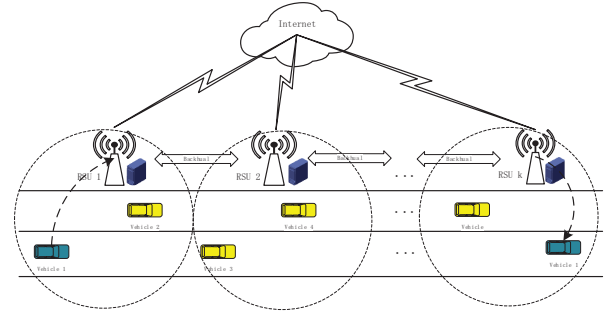}\caption{The task offloading via I2I migration link.\label{fig:The-offloading-via-1} }
\end{figure}

For V2I offloading, we can obtain the delay $T_{m,\mathrm{mig}}^{\mathrm{I2I,}}$,
and the energy consumption $E_{m,\mathrm{mig}}^{\mathrm{I2I}}(\boldsymbol{d})$, given by
\begin{align}
 & T_{\mathrm{\mathit{m},mig}}^{\mathrm{I2I}}(\boldsymbol{d})\nonumber \\
 & =T_{m}^{\mathrm{V2I}}(\boldsymbol{d})+\varphi\frac{D_{m,\mathrm{out}}}{R_{\mathrm{\mathit{m},mig}}^{\mathrm{I2I}}(\boldsymbol{d})}\nonumber \\
 & =\frac{D_{\mathrm{\mathit{m},out}}+D_{\mathrm{\mathit{m},in}}}{R_{\mathrm{\mathit{m},tran}}^{\mathrm{V2I}}(\boldsymbol{d})}+\frac{C_{m,\mathrm{in}}}{\mathit{f}_{\mathrm{mec}}}+\varphi\frac{D_{m,\mathrm{out}}}{R_{m,\mathrm{mig}}^{\mathrm{I2I}}(\boldsymbol{d})},\label{eq:I2I_delay}
\end{align}
\begin{align}
 & E_{\mathrm{\mathit{m},mig}}^{\mathrm{I2I}}(\boldsymbol{d})\nonumber \\
 & =p_{\mathrm{\mathit{m},tran}}^{\mathrm{V2I}}\frac{D_{m,\mathrm{in}}}{R_{\mathrm{\mathit{m},tran}}^{\mathrm{V2I}}(\boldsymbol{d})}+p_{\mathrm{\mathit{m},tran}}^{\mathrm{V2I}}\frac{D_{m,\mathrm{out}}}{R_{\mathrm{\mathit{m},tran}}^{\mathrm{V2I}}(\boldsymbol{d})},\label{eq:I2I_energy}
\end{align}
where $\varphi$ denotes the number of I2I migration RSUs when the target vehicle receives the computing result. In the system,
we consider only the energy consumption of vehicles: the energy
consumption of transmission between RSUs is not taken into account.
Fig. \ref{fig:The-offloading-via-1} clearly illustrates
the process of I2I task migration. As in V2V task migration,
when a vehicle processes a task in I2I migration mode, the vehicle
offloads the task to a nearby RSU through the V2I communication
link. The RSU processes the task and then transmits the computing
result to an RSU that has the target vehicle in its coverage area via
the I2I migration link. After migration, the RSU transmits the
computing result to the target vehicle.

{\begin{problem}
When the target vehicle received the computing result, how to compute the value of $\varphi$ that the number of I2I migrations that the target vehicle passed?
\end{problem}
\begin{sol}
As shown in Fig. \ref{fig:The-offloading-via-1}, we know that the distance of the target vehicle driven is greater than $\varphi$ times the average distance between RSUs and less than $\varphi$ times the average distance between RSUs plus the coverage diameter of RSU within the time when the target vehicle obtains the computing results. Therefore, we can obtain
\[
\varphi L_{\mathrm{I2I}}<\bar{\upsilon}T_{m,\mathrm{mig}}^{\mathrm{I2I}}(\boldsymbol{d})<\varphi L_{\mathrm{I2I}}+L_{\mathrm{I2I}},
\]
\[
\Rightarrow\begin{cases}
\varphi L_{\mathrm{I2I}}<\bar{\upsilon}(T_{m}^{\mathrm{V2I}}(\boldsymbol{d})+\varphi\frac{D_{m,\mathrm{out}}}{R_{m,\mathrm{mig}}^{\mathrm{I2I}}(\boldsymbol{d})}),\\
\bar{\upsilon}(T_{m}^{\mathrm{V2I}}(\boldsymbol{d})+\varphi\frac{D_{m,\mathrm{out}}}{R_{m,\mathrm{mig}}^{\mathrm{I2I}}(\boldsymbol{d})})<(\varphi+1)L_{\mathrm{I2I}},
\end{cases}
\]
\[
\Rightarrow\begin{cases}
\varphi>\frac{\bar{\upsilon}(\frac{D_{m,\mathrm{in}}}{R_{m,\mathrm{tran}}^{\mathrm{V2I}}(\boldsymbol{d})}+\frac{C_{m,in}}{\mathit{f}_{\mathrm{mec}}}+\frac{D_{m,\mathrm{out}}}{R_{m,\mathrm{tran}}^{\mathrm{V2I}}(\boldsymbol{d})})-L_{\mathrm{I2I}}}{L_{\mathrm{I2I}}-\bar{\upsilon}\frac{D_{m,\mathrm{out}}}{R_{m,\mathrm{mig}}^{\mathrm{I2I}}(\boldsymbol{d})}},\\
\varphi<\frac{\bar{\upsilon}(\frac{D_{m\mathrm{,in}}}{R_{m,\mathrm{tran}}^{\mathrm{V2I}}(\boldsymbol{d})}+\frac{C_{m,\mathrm{in}}}{\mathit{f}_{\mathrm{mec}}}+\frac{D_{m,\mathrm{out}}}{R_{m,\mathrm{tran}}^{\mathrm{V2I}}(\boldsymbol{d})})}{L_{\mathrm{I2I}}-\bar{\upsilon}\frac{D_{m,\mathrm{out}}}{R_{m,\mathrm{mig}}^{\mathrm{I2I}}(\boldsymbol{d})}},
\end{cases}
\]
subject to
\begin{align*}
 & L_{\mathrm{I2I}}<\bar{\upsilon}(\frac{D_{m,\mathrm{in}}}{R_{m,\mathrm{tran}}^{\mathrm{V2I}}(\boldsymbol{d})}+\frac{C_{m,\mathrm{in}}}{\mathit{f}_{\mathrm{mec}}}+\frac{D_{m,\mathrm{out}}}{R_{m,\mathrm{tran}}^{\mathrm{V2I}}(\boldsymbol{d})}),\\
 & L_{\mathrm{I2I}}>\bar{\upsilon}\frac{D_{m,\mathrm{out}}}{R_{m,\mathrm{mig}}^{\mathrm{I2I}}(\boldsymbol{d})}.
\end{align*}

In the system, we express the average of $\bar{\varphi}$ as
\[
\bar{\varphi}=\lfloor\frac{2\bar{\upsilon}(\frac{D_{m,\mathrm{in}}}{R_{m,\mathrm{tran}}^{\mathrm{V2I}}(\boldsymbol{d})}+\frac{C_{m,\mathrm{in}}}{\mathit{f}_{\mathrm{mec}}}+\frac{D_{m,\mathrm{out}}}{R_{m,\mathrm{tran}}^{\mathrm{V2I}}(\boldsymbol{d})})-L_{\mathrm{I2I}}}{2(L_{\mathrm{I2I}}-\bar{\upsilon}\frac{D_{m,\mathrm{out}}}{R_{m,\mathrm{mig}}^{\mathrm{I2I}}(\boldsymbol{d})})}\rfloor,
\]where $\lfloor.\rfloor$ denotes rounding down.

According to \eqref{eq:local_computation_overhead}, we can obtain computation overhead $\Omega_{m,\mathrm{mig}}^{\mathrm{I2I}}(\boldsymbol{d})$
of migrating computing results via the I2I migration link as
\begin{equation}
\Omega_{m,\mathrm{mig}}^{\mathrm{I2I}}(\boldsymbol{d})=\alpha_{m}T_{m,\mathrm{mig}}^{\mathrm{I2I}}(\boldsymbol{d})+\beta_{m}E_{m,\mathrm{mig}}^{\mathrm{I2I}}(\boldsymbol{d}).\label{eq:I2I_computation_overhead}
\end{equation}
\end{sol}

\subsection{Problem Formulation}

In this section, according to the task offloading decisions $\mathit{\boldsymbol{d}_{-m}}\stackrel{\bigtriangleup}{=}\left(d_{1},...,d_{m-1},d_{m+1},...,d_{M}\right)$
of all other vehicles except vehicle $m$, we analyze the impact of
the offloading decision of vehicle $m$ on the overall computation
overhead.

\subsubsection{Minimize the Computation Overhead of the System}

In this section, we discuss how to minimize the computation overhead
of the whole network. According to the discussion in Section \mbox{III},
vehicle $m$ has three alternative ways to process its task and
two alternative ways to migrate the computing result. The choice
of vehicle $m$ directly affects the computation overhead of the whole
network. As the number
of tasks in the V2V communication link, V2I communication link, V2V
migration link and I2I migration link increases, their transmission rates decrease,
thereby increasing the transmission delay and energy consumption. As a result,
the computation overhead of the whole network increases, which
is contrary to our purpose. In this paper, we design an offloading
scheme under various constraints to minimize the computation overhead
$\Upsilon(\boldsymbol{d})$ of the whole system. Under the given tasks offloading decisions $\boldsymbol{d}=\{d_{1},d_{2},d_{3},...,d_{M}\}$
of all other vehicles, we can obtain the computation overhead $\Upsilon(\boldsymbol{d})$
of the network which is described as a mathematical problem, as given
by
\begin{align}
\mathbf{P1}: & \underset{\boldsymbol{d},\tau,\alpha_{m},\beta_{m},\mu_{m}}{\min}\Upsilon(\boldsymbol{d})=\sum_{m=1}^{M}(\Gamma_{\{d_{m}=0\}}\Omega_{m}^{\mathrm{local}}\nonumber \\
 & +\Gamma_{\{d_{m}=1\}}\Omega_{m}^{\mathrm{V2V}}+\Gamma_{\{d_{m}=2\}}\Omega_{m}^{\mathrm{V2I}}\nonumber \\
 & +\Gamma_{\{d_{m}=3\}}\Omega_{m,\mathrm{mig}}^{\mathrm{V2V}}+\Gamma_{\{d_{m}=4\}}\Omega_{m,\mathrm{mig}}^{\mathrm{I2I}}),\label{eq:question}
\end{align}
\begin{eqnarray*}
s.t. & C1: & 0\leq p_{\mathrm{\mathit{m},tran}}^{\mathrm{V2V}}\leq p_{\mathrm{\mathit{m},tran}}^{\mathrm{V2V},\max},\forall m\in\mathbf{M},\\
 & C2: & 0\leq p_{\mathrm{\mathit{m},tran}}^{\mathrm{V2I}}\leq p_{\mathrm{\mathit{m},tran}}^{\mathrm{V2I},\max},\forall m\in\mathbf{M},\\
 & C3: & 0\leq p_{\mathrm{\mathit{m},mig}}^{\mathrm{V2V}}\leq p_{\mathrm{\mathit{m},mig}}^{\mathrm{V2V},\max},\forall m\in\mathbf{M},\\
 & C4: & 0\leq p_{\mathrm{\mathit{m},mig}}^{\mathrm{I2I}}\leq p_{\mathrm{\mathit{m},mig}}^{\mathrm{V2V},\max},\forall m\in\mathbf{M},\\
 & C5: & \sum_{m=1}^{M}\Gamma_{\{d_{m}=g\}}\leq M,\forall m\in\mathbf{M},g\in\{0,1,2,3,4\},\\
 & C6: & d_{m}\in\Psi,\forall m\in\mathbf{M},\\
 & C7: & T_{m}^{\mathrm{local}}\leq\tau,\forall m\in\mathbf{M},\\
 & C8: & T_{m}^{\mathrm{V2V}}(\boldsymbol{d})\leq\tau,\forall m\in\mathbf{M},\\
 & C9: & T_{m}^{\mathrm{V2I}}(\boldsymbol{d})\leq\tau,\forall m\in\mathbf{M},\\
 & C10: & T_{m,\mathrm{mig}}^{\mathrm{V2V}}(\boldsymbol{d})\leq\tau,\forall m\in\mathbf{M},\\
 & C11: & T_{m,\mathrm{mig}}^{\mathrm{I2I}}(\boldsymbol{d})\leq\tau,\forall m\in\mathbf{M},\\
 & C12: & \alpha_{m}+\beta_{m}=1,\forall m\in\mathbf{M},\\
 & C13: & \mu_{m}\in(0,1],
\end{eqnarray*}
where $\Gamma_{\{d_{m}=g\}}$is an indicator function. If $d_{m}=g$
is true, $\Gamma_{\{d_{m}=g\}}=1$; otherwise, $\Gamma_{\{d_{m}=g\}}=0$.

In P1, C1, C2, C3 and C4 are the maximum transmission power constraints
of the V2V communication link, V2I communication link, V2V migration link
and I2I migration link, respectively. C5 denotes the constraint of
the number of vehicles that use five modes to process the computing
task. C6 denotes the offloading decision choice of vehicle $m$.
C7, C8, C9, C10 and C11 are the maximum transmission delay constraints
of local processing, V2V processing, V2I processing, V2V migrating
and I2I migrating, respectively. C12 indicates the relationship between
the weight of delay and the weight of energy consumption. C13 denotes
the constraint of the available rate of computational resources.

When the target vehicle uses task migration, the following constraint settings
must be met, as given by
\begin{eqnarray*}
s.t. & S1: & L_{\mathrm{I2I}}<\bar{\upsilon}(\frac{D_{m,\mathrm{in}}}{R_{m,\mathrm{tran}}^{\mathrm{V2I}}(\boldsymbol{d})}+\frac{C_{m,\mathrm{in}}}{\mathit{f}_{\mathrm{mec}}}+\frac{D_{m,\mathrm{out}}}{R_{m,\mathrm{tran}}^{\mathrm{V2I}}(\boldsymbol{d})}),\\
 &  & \forall m\in\mathbf{M},\\
 & S2: & L_{\mathrm{V2V}}>\bar{\upsilon}\frac{D_{m,\mathrm{out}}}{R_{m,\mathrm{mig}}^{\mathrm{V2V}}(\boldsymbol{d})},\forall m\in\mathbf{M},\\
 & S3: & L_{\mathrm{I2I}}>\bar{\upsilon}\frac{D_{m,\mathrm{out}}}{R_{m,\mathrm{mig}}^{\mathrm{I2I}}(\boldsymbol{d})},\forall m\in\mathbf{M},
\end{eqnarray*}
where S1 indicates that the vehicle must be driven out of RSU communication
range within a limited time to use the migration. S2 indicates
that when a vehicle adopts V2V migration to transmit the computing
result, the distance that the vehicle drives in a single V2V migration
time should be less than the average distance between two vehicles; otherwise, the vehicle will never catch up with the computing result.
Similarly, S3 indicates that when a vehicle adopts I2I migration
to transmit the computing result, the distance that the vehicle drives
in a single I2I migration time should be less than the average distance
between two RSUs; otherwise, the vehicle will never catch up with the
computing result.

In the following sections, we provide a game theory-based strategy to
solve the problem of offloading decisions among vehicles. This strategy can effectively
reduce the weighted sum of the delay and energy consumption of the system.

\subsubsection{Minimize the Computation Overhead of the Vehicle}

In this section, we discuss how to minimize the computation overhead
of the vehicle. The computation offloading decisions
$\boldsymbol{d}$ of the vehicles are coupled in the system. When
a large number of vehicles choose the same link to offload tasks or
to migrate the computing results, they can cause interference between
tasks and slow the transmission and migration rates. Even very low
transmission rates and migration rates can increase transmission
delays, transmission energy consumption, migration delays and migration
energy consumption. In this case, it is practical and efficient for
the vehicles to choose local processing of computation tasks with task
migration to avoid the long processing delay and high energy consumption.
V2V migration and I2I migration can be reasonably used for tasks requiring
computing migration.

Under the given task offloading decisions $\mathit{\boldsymbol{d}_{-m}}\stackrel{\bigtriangleup}{=}\left(d_{1},...,d_{m-1},d_{m+1},...,d_{M}\right)$
of all other vehicles except vehicle $m$, we obtain the computation
overhead function of vehicle $m$ as
\begin{equation}
\Omega_{m}(d_{m},\boldsymbol{d}_{-m})=\begin{cases}
\Omega_{m}^{\mathrm{local}}, & if\;d_{m}=0,\\
\Omega_{m}^{\mathrm{V2V}}, & if\;d_{m}=1,\\
\Omega_{m}^{\mathrm{V2I}}, & if\;d_{m}=2,\\
\Omega_{m,\mathrm{mig}}^{\mathrm{V2V}}, & if\;d_{m}=3,\\
\Omega_{m,\mathrm{mig}}^{\mathrm{I2I}}, & if\;d_{m}=4.
\end{cases}
\end{equation}

In the following sections, we provide a game theory-based strategy to solve
the problem of offloading decisions among vehicles to effectively
reduce the weighted sum of the delay and energy consumption of the vehicle.

\section{Computation Offloading and Task Migration}

In this section, we provide a game approach for the considered computation
offloading system, and then formulate the offloading decision-making
problem as a game.

Vehicles belong to rational (or selfish) people whose goal is to prevent damage to their own interests, and game theory considers how selfish players make decisions. Thus, we choose game theory as our solution. In game theory, because people are rational
(or selfish), each person chooses the decision that minimizes his
or her own computation overhead, regardless of the positive or negative impact
of his or her decision on others. Game theory is an effective method
for vehicles to analyze their interests and obtain the minimum computation
overhead through offloading decisions.

\subsection{Game Formulation}

Different vehicles have different interests and capabilities, and run different
applications. Analyzing the interactions among multiple vehicles by means of
game theory is convenient. Specifically, we consider that vehicles
choose their offloading decisions sequentially. We formulate the
offloading decision-making problem as a game. Game theory
has the advantage of not only helping to develop low-complexity algorithms,
but also reducing the amount of computation required
by algorithms.

By changing the offloading decision of vehicle $m$, we can obtain
the minimized weighted sum of delay and energy consumption
of all vehicles within a finite number of iterations. The weighted
sum of delay and energy consumption is related not only to the offloading
strategy of vehicle $m$ but also to the offloading decisions
of other vehicles. This is because when the offloading decision of vehicle
$m$ changes, the transmission rate or migration rate of other vehicles
may change, thus affecting the computation overhead of other vehicles.
Given offloading decisions $\mathit{\boldsymbol{d}_{-m}}\stackrel{\bigtriangleup}{=}\left(d_{1},...,d_{m-1},d_{m+1},...,d_{M}\right)$
of other vehicles, vehicle $m$ can choose the best offloading
decision $d_{m}$ to minimize its own computation overhead, i.e.,
\[
\underset{d_{m}\in\{0,1,2,3,4\}}{\min}\Omega_{m}(d_{m},\boldsymbol{d}_{-m}),\:\forall m\in\mathbf{M}.
\]

When the offloading decisions of other vehicles are known, vehicle
$m$ can select the offloading decision with the lowest computation
overhead by comparing the computation overhead under the five offloading
decisions. Then, we describe the above problem as the offloading decision
game $\Delta=(\mathbf{M},\mathbf{\Psi},\{\Omega_{m}\}_{m\in\mathbf{M}})$,
where $\mathbf{M}$ is the finite set of the player, $\mathbf{\Psi}$
is the set of strategies for vehicle $m$, and $\left\{ \Omega_{m}\right\} _{m\in\mathbf{M}}$
is the set of computation overhead functions of vehicle $m$. In game theory, the NE is an important point. Next, we will introduce the NE.
\begin{thm}
If only one player that has finite action sets makes a decision at a time and knows every action of other players, we call the game $\Delta$ a perfect information sequential offloading game. A finite sequential game with perfect information exists an NE, and it can be converged within a finite number of
decision slots, which is called the FIP.
\end{thm}
\begin{IEEEproof}
The proof is given in \cite{47}.
\end{IEEEproof}
\begin{defn}
For the sequential offloading game $\Delta$, if no player can change
his/her offloading decision to reduce the overall computation overhead
under given offloading decision $\mathit{\boldsymbol{d}_{-m}}\stackrel{\bigtriangleup}{=}\left(d_{1},...,d_{m-1},d_{m+1},...,d_{M}\right)$,
the offloading decision $\boldsymbol{d}^{*}$ is the Nash equilibrium
of the game if and only if the following conditions are satisfied:
\[
\Omega_{m}(d_{m}^{*},\boldsymbol{d}_{-m}^{*})\leq\Omega_{m}(d_{m},\boldsymbol{d}_{-m}^{*}),\:\forall m\in\mathbf{M},\:\forall d_{m}\in\mathbf{\Psi}.
\]

According to the FIP, after a certain number of iterations, the game
theory will move towards the NE which is stable. It
is means that, under the given set of offloading decisions $\mathit{\boldsymbol{d}_{-m}}\stackrel{\bigtriangleup}{=}\left(d_{1},...,d_{m-1},d_{m+1},...,d_{M}\right)$
of other vehicles, vehicle $m$ cannot reduce the total computation
overhead of all vehicles by changing its own offloading decision.
Therefore, based on the Nash equilibrium principle of game theory,
each vehicle can obtain its own satisfactory offloading decision.
\end{defn}

\subsection{Convergence}

In this section, we will discuss the convergence of game theory. When
a finite number of players participate in a game, and each player
has a finite number of decisions to choose, the game can reach NE in a finite number of steps \cite{35,36}, according to the existence
theorem of NE.
\begin{thm}
Starting at any offloading decision, the offloading
decision-making game of Algorithm \ref{alg:Computing-offloading-algorithm}
is guaranteed to eventually converge.
\end{thm}
\begin{IEEEproof}
See Appendix \ref{prf1}.
\end{IEEEproof}

\subsection{Complexity}
Each cycle performs $M$ iterations to calculate the
minimum computation overhead per vehicle. When Algorithm \ref{alg:Computing-offloading-algorithm}
performs $Q$ cycles to reach NE, the complexity
of the algorithm is $O(MQ)$, which is much lower than the
complexity of the exhaustive array of $O(5^{M})$.

\subsection{Task Migration}

In this section, we consider the case of using task migration when the amount of computations is large  and the speed is fast.

In the MEC-assisted vehicular network, the RSU has two links to transmit
the computing result, i.e., the V2V migration link and the I2I migration link.
According to the discussion in Section \mbox{III}, considering the
co-channel interference, when too many tasks use
the same migration link to transmit the computing results, the migration
rate will be slow, and the migration delay and migration energy
consumption will increase. Therefore, in this subsection, we make reasonable use
of migration links and transmission links to minimize the vehicle
computation overhead. Next, we introduce the migration selection
algorithm and the specific steps of the migration selection algorithm, as shown in Algorithm \ref{alg:The-Migration-Selection}.
\begin{algorithm}
\caption{Migration Selection (MS) Algorithm \label{alg:The-Migration-Selection}}

\begin{enumerate}
\item $\mathbf{Input}$: $\mathrm{\mathit{f}_{ue}}$, $\mu_{m}$, $k$,
$\mathit{f_{\mathrm{mec}}}$. Task ($D_{m,\mathrm{in}},C_{m,\mathrm{in}},\tau)$,
$\alpha_{m}$, $\beta_{m}$, vehicle $m\{m=1,2,3...M\}$.
\item $\mathbf{\mathbf{Outpu}t}$: offloading decisions $\boldsymbol{d}=(d_{1},d_{2},d_{3},...,d_{M})$
and the computation overhead $\varUpsilon(\boldsymbol{d})$.
\item $\mathbf{initialization}$: offloading decisions of all vehicles are
set as 0, i.e., $d_{m}=0,\:\forall m\in\mathbf{M}$, so the set of
offloading decisions is $\boldsymbol{d}=(0_{1},0_{2},0_{3},...,0_{M})$
in the beginning.
\item $\mathbf{begin}$:
\item $\qquad$vehicle $m$ generates computing task
\item[] $\qquad$ $(D_{m,\mathrm{in}},C_{m,\mathrm{in}},\tau)$;
\item $\qquad$vehicle $m$ offloads task to the nearby RSU $s$;
\item $\qquad$the RSU $s$ processes the task;
\item $\qquad$$\mathbf{if}$ vehicle $m$ is still in the communication
range
\item[] $\qquad$of RSU $s$ when RSU $s$ has processed the
    \item[] $\qquad$task $\mathbf{then}$
\item $\qquad\qquad$let vehicle $m$ receive the computing results
\item[] $\qquad\qquad$directly from RSU $s$ via a V2I \item[] $\qquad\qquad$    communication link;
\item $\qquad\qquad$calculate the computation overhead $\Omega_{m}^{\mathrm{V2I}}(\boldsymbol{d})$
\item[] $\qquad\qquad$of vehicle $m$;
\item $\qquad$$\mathbf{else}$
\item $\qquad\qquad$let the computing results be transmitted via a
    \item[] $\qquad\qquad$V2V migration
link or an I2I migration link;
\item $\qquad\qquad$after migrating, let vehicle $m$ receive the
    \item[] $\qquad\qquad$computing
result eventually;
\item $\qquad\qquad$calculate the computation overhead
\item[] $\qquad\qquad$ $\Omega_{m,\mathrm{mig}}^{\mathrm{V2V}}(\boldsymbol{d})/\Omega_{m,\mathrm{mig}}^{\mathrm{I2I}}(\boldsymbol{d})$
of vehicle $m$;
\item $\qquad$$\mathbf{end\:if}$
\item $\mathbf{end}$
\end{enumerate}
\end{algorithm}

\subsection{Computation Overhead Minimization Offloading Algorithm}

Based on the above study, we propose a task migration (TM) algorithm
based on game theory and Algorithm \ref{alg:The-Migration-Selection}.
In the TM algorithm, the offloading decision with the lowest computation
overhead is determined by changing the offloading decision of the
vehicle. Using the above approach, after a period of iteration,
the game will reach NE, and each vehicle makes the offloading decision that minimizes its own computation overhead.
Considering the task offload situation, for each vehicle, the TM algorithm
achieves the minimum value of its own computation overhead on the
condition that the offloading decisions of other vehicles are known.
The detailed procedure of the proposed algorithm is shown in Algorithm
\ref{alg:Computing-offloading-algorithm}.
\begin{algorithm}[h]
\caption{Task Migration (TM) Algorithm \label{alg:Computing-offloading-algorithm} }

\begin{enumerate}
\item $\mathbf{Input}$: $\mathrm{\mathit{f}_{ue}}$, $\mu_{m}$, $k$,
$\mathit{f_{\mathrm{mec}}}$. Task ($D_{m,\mathrm{in}},C_{m,\mathrm{in}},\tau)$,
$\alpha_{m}$, $\beta_{m}$, vehicle $m\{m=1,2,3...M\}$.
\item $\mathbf{\mathbf{Outpu}t}$: offloading decisions $\boldsymbol{d}=(d_{1},d_{2},d_{3},...,d_{M})$
and the computation overhead $\varUpsilon(\boldsymbol{d})$.
\item $\mathbf{initialization}$: offloading decisions of all vehicles are
set as 0, i.e., $d_{m}=0,\:\forall m\in\mathbf{M}$, so the set of
offloading decisions is $\boldsymbol{d}=(0_{1},0_{2},0_{3},...,0_{M})$
in the beginning.
\item $\mathbf{begin}:$
\item let vehicle $m$ generate a computing task;
\item $\mathbf{repeat}$
\item $\mathbf{for}$ $m\in\mathbf{M}$ $\mathrm{\mathbf{do}}$
\item $\qquad$calculate the rates based on known offloading
\item[] $\qquad$    decision $\boldsymbol{d}$;
\item $\qquad$calculate $T_{m}^{\mathrm{local}}$, $T_{m}^{\mathrm{V2V}}(\boldsymbol{d})$,
$T_{m}^{\mathrm{V2I}}(\boldsymbol{d})$, $T_{m,\mathrm{mig}}^{\mathrm{V2V}}(\boldsymbol{d})$
\item[] $\qquad$and $T_{m,\mathrm{mig}}^{\mathrm{I2I}}(\boldsymbol{d})$ according
to \eqref{eq:local_delay}, \eqref{eq:V2V_delay}, \eqref{eq:V2I_delay},
\eqref{eq:V2V_migration_delay} and
\item[] $\qquad$\eqref{eq:I2I_delay};
\item $\boldsymbol{\qquad\mathbf{if}}$ all the delays are greater than
$\tau$ $\mathbf{then}$
\item $\qquad\qquad$the task cannot be completed on time;
\item $\qquad$$\boldsymbol{\mathbf{else}}$
\item $\qquad\qquad$calculate $\Omega_{m}^{\mathrm{local}}$, $\Omega_{m}^{\mathrm{V2V}}(\boldsymbol{d})$,
$\Omega_{m}^{\mathrm{V2I}}(\boldsymbol{d})$,
\item[] $\qquad\qquad$
$\Omega_{m,\mathrm{mig}}^{\mathrm{V2V}}(\boldsymbol{d})$
and $\Omega_{m,\mathrm{mig}}^{\mathrm{I2I}}(\boldsymbol{d})$ according
to \eqref{eq:local_computation_overhead},
\item[] $\qquad\qquad$ \eqref{eq:V2V_computation_overhead},
\eqref{eq:V2I_computation_overhead}, \eqref{eq:V2V_migration_computation_overhead}
and \eqref{eq:I2I_computation_overhead};
\item $\qquad\qquad$take the minimum of $\Omega_{m}^{\mathrm{local}}$,
$\Omega_{m}^{\mathrm{V2V}}(\boldsymbol{d})$,
\item[] $\qquad\qquad$ $\Omega_{m}^{\mathrm{V2I}}(\boldsymbol{d})$,
$\Omega_{m,\mathrm{mig}}^{\mathrm{V2V}}(\boldsymbol{d})$ and $\Omega_{m,\mathrm{mig}}^{\mathrm{I2I}}(\boldsymbol{d})$
whose
\item[] $\qquad\qquad$delay is less than $\tau$;
\item $\qquad\qquad$calculate $d_{m}$, which is the minimum value;
\item $\qquad$$\mathbf{end\:if}$
\item $\mathbf{end\:for}$
\item obtain the offloading decision $\boldsymbol{d^{*}}$;
\item $\mathbf{if}$ $\boldsymbol{d\neq d^{*}}$ $\mathbf{then}$
\item $\qquad$send update-request (UR) to the server;
\item $\qquad$$\mathbf{if}$ receive the permission of UR $\mathbf{then}$
\item $\qquad\qquad$obtain the offloading decision $\boldsymbol{d^{*}}$;
\item $\qquad\qquad$update the offloading decision $\boldsymbol{d}=\boldsymbol{d}^{*}$;
\item $\qquad$$\mathbf{end\:if}$
\item $\mathbf{end\:if}$
\item $\mathbf{until}$ no UR for the offloading decision.
\item calculate the computation overhead $\varUpsilon(\boldsymbol{d})$;
\item $\mathbf{return}$ offloading decision $\boldsymbol{d}$ and computation
overhead $\varUpsilon(\boldsymbol{d})$.
\item $\mathbf{end}$
\end{enumerate}
\end{algorithm}

In Algorithm \ref{alg:Computing-offloading-algorithm}, according to game theory, each vehicle chooses the offloading
decision that minimizes its own computation overhead. After a finite
number of iterations, the game reaches NE,
and the vehicle reaches the offloading decision that minimizes its
own computation overhead under the condition that the offloading decisions
of other vehicles are known. In this equilibrium state, the computation
overhead for each vehicle is minimal, but the computation overhead
for the whole system may not reach the minimum value. Therefore, the offloading
decisions obtained by game theory may not be optimal for the computation
overhead of the system. To minimize the computation overhead of the
whole system, we propose a computation overhead minimization offloading
(COMO) algorithm, whose detailed steps are described in Algorithm \ref{alg:Optimal-Offloading-(OO)}.
\begin{algorithm}
\caption{Computation Overhead Minimization Offloading (COMO) Algorithm \label{alg:Optimal-Offloading-(OO)}}

\begin{enumerate}
\item $\mathbf{Input}$: $\mathit{f_{\mathrm{ue}}}$, $k$, $\mu_{m}$,
$\mathrm{\mathit{f_{\mathrm{mec}}}}$, $(D_{m,\mathrm{in}},C_{m,\mathrm{in}},\tau)$,
$\alpha_{m}$, $\beta_{m}$, vehicle $m\{m=1,2,3...M\}$.
\item $\mathbf{\mathbf{Outpu}t}$: offloading decision $\boldsymbol{d}=(d_{1},d_{2},d_{3},...,d_{M})$
, computation overhead $\varUpsilon(\boldsymbol{d})$ .
\item $\mathbf{initialization}$: each vehicle $m$ chooses the offloading
decision $d_{m}=0,\:\forall m\in\mathbf{M}$.
\item $\mathbf{begin}$
\item let vehicle $m$ generate a computing task;
\item $\mathbf{repeat}$
\item $\mathbf{for}$ $m\in\mathbf{M}$ $\mathrm{\mathbf{do}}$
\item $\qquad$obtain offloading decision $\boldsymbol{d}$;
\item $\qquad$$\mathbf{for}$ each time slot $t$;
\item $\qquad\qquad$offloading decision $d_{m,t}\rightarrow d_{m,t+1}$;
\item $\qquad\qquad$computation overhead $\varUpsilon(\boldsymbol{d}_{t})$$\rightarrow\varUpsilon(\boldsymbol{d}_{t+1})$;
\item $\qquad\qquad$obtain $\boldsymbol{d}_{t+1}^{*}$;
\item $\qquad\qquad$$\mathbf{if}$ $\varUpsilon(\boldsymbol{d}_{t})<\varUpsilon(\boldsymbol{d}_{t+1})$
$\mathbf{then}$
\item $\qquad\qquad\qquad$$\boldsymbol{d}_{t+1}^{*}\boldsymbol{=d}_{t+1}$$^{*}$;
\item $\qquad\qquad$$\mathbf{else}$
\item $\qquad\qquad\qquad$ $\boldsymbol{d}_{t+1}^{*}$$\neq$$\boldsymbol{d}_{t}^{*}$;
\item $\qquad\qquad$$\mathbf{end\;if}$
\item $\qquad$$\mathbf{end\;for}$
\item $\qquad$obtain the offloading decision $\boldsymbol{d}$$^{*}$;
\item $\qquad$$\mathbf{if}$ $\boldsymbol{d\neq d}$$^{*}$ $\mathbf{then}$
\item $\qquad\qquad$the server asks the vehicles to update their
    \item[] $\qquad\qquad$offloading decisions;
\item $\qquad\qquad$update the offloading decision $\boldsymbol{d=d}$$^{*}$;
\item $\qquad$$\mathbf{end\:if}$
\item $\mathbf{end\:for}$
\item $\mathbf{until}$ the server does not send an update request.
\item $\mathbf{return}$ offloading decision $\boldsymbol{d}$ and computation
overhead $\varUpsilon(\boldsymbol{d})$.
\end{enumerate}
\end{algorithm}

\begin{table*}
\caption{SIMULATION PARAMETERS}

\centering%
\begin{tabular}{|c|c||c|c|}
\hline
Parameter & Value & Parameter & Value\tabularnewline
\hline
\hline
Distance between adjacent vehicles $L_{\mathrm{V2V}}$ & $100\:m$ & The available rate of computational resources $\mu$ & $0.5\sim1$\tabularnewline
\hline
Distance between adjacent RSUs $L_{\mathrm{I2I}}$  & $300\:m$ & The power of complex Gaussian white noise $\omega$ & $10^{-10}$\tabularnewline
\hline
Distance between RSU and vehicle $L_{\mathrm{V2I}}$ & $100\:m$ & The transmission power of the V2I communication link $p_{\mathrm{\mathit{m},tran}}^{\mathrm{V2I}}$ & $\mathrm{400\:mW}$\tabularnewline
\hline
The size of data to be processed $D_{m,\mathrm{in}}$ & $5\mathrm{\mathrm{\:Mbit\sim50\:Mbi}t}$ & The bandwidth of the V2I communication link $\mathrm{b_{\mathit{m},tran}^{V2I}}$  & $\mathrm{10\:MHz}$\tabularnewline
\hline
The number of cycles by the CPU $C_{n,\mathrm{in}}$ & $\delta D_{n}\:cycles$ & The channel gain of the V2I communication link $h_{\mathrm{\mathit{m},tran}}^{\mathrm{V2I}}$ & $L_{\mathrm{V2I}}^{-\theta}$\tabularnewline
\hline
$\delta$ & $3.055*10^{3}$ & The bandwidth of the V2V migration link $\mathrm{b_{\mathit{m},mig}^{V2V}}$  & $\mathrm{4\:MHz}$\tabularnewline
\hline
The size of the feedback data $D_{m,\mathrm{out}}$ & $D_{m,\mathrm{in}}=\xi D_{m,\mathrm{out}}$ & The transmission power of the V2V migration link $p_{\mathrm{\mathit{m},mig}}^{\mathrm{V2V}}$ & $\mathrm{100\:mW}$\tabularnewline
\hline
$\xi$ & $100$ & The channel gain of the V2V migration link $h_{\mathrm{\mathit{m},mig}}^{\mathrm{V2V}}$ & $L_{\mathrm{V2V}}^{-\theta}$\tabularnewline
\hline
The CPU cycle frequency of vehicle $\mathit{f}_{\mathrm{ue}}$ & $\mathrm{1\:GHz}$ & The bandwidth of the I2I migration link $\mathrm{b_{\mathit{m},mig}^{I2I}}$  & $\mathrm{4\:MHz}$\tabularnewline
\hline
The CPU cycle frequency of RSU $\mathit{f}_{\mathrm{mec}}$ & $\mathrm{5\:GHz}$ & The transmission power of the I2I migration link $p_{\mathrm{\mathit{m},mig}}^{\mathrm{I2I}}$ & $\mathrm{400\:mW}$\tabularnewline
\hline
The energy coefficient $k$  & $10^{-28}$ \cite{26} & The channel gain of the I2I migration link $h_{\mathrm{\mathit{m},mig}}^{\mathrm{I2I}}$ & $L_{\mathrm{I2I}}^{-\theta}$\tabularnewline
\hline
The weights of delay $\alpha_{m}$ & $0\sim1$ & The transmission power of the V2V communication link $p_{\mathrm{\mathit{m},tran}}^{\mathrm{V2V}}$ & $\mathrm{100\:mW}$\tabularnewline
\hline
The weights of energy consumption $\beta_{m}$ & $0\sim1$ & The bandwidth of the V2V communication link $\mathrm{b_{\mathit{m},tran}^{V2V}}$  & $\mathrm{10\:MHz}$\tabularnewline
\hline
The path loss exponent $\theta$ & $4$ & The channel gain of the V2V communication link $h_{\mathrm{\mathit{m},tran}}^{\mathrm{V2V}}$ & $L_{\mathrm{V2V}}^{-\theta}$\tabularnewline
\hline
The average speed $\bar{\upsilon}$ & $108\:\mathrm{km/h}$ & The average number of vehicles that pass by V2V migration $\bar{\phi}$ & \tabularnewline
\hline
The maximum delay tolerance for the task $\tau$ &  & The average number of RSUs that pass by I2I migration $\bar{\varphi}$ & \tabularnewline
\hline
\end{tabular}\label{Table1}
\end{table*}

In Algorithm \ref{alg:Optimal-Offloading-(OO)}, the best offloading direction is similar to the steepest descent direction in reducing the computation overhead of the whole system. If $\varUpsilon(\boldsymbol{d}_{t})$$\nless\varUpsilon(\boldsymbol{d}_{t+1})$,
the offloading decision of the vehicle will be updated in the next
time slot; otherwise, the vehicle will keep the offloading decision
unchanged in the next time slot, i.e., $d_{m,t+1}=d_{m,t}$. For
the offloading decision updating problem, we set the number of vehicles
as the number of iterations of each cycle. When the offloading decision
of each vehicle does not change at the end of each cycle iteration
(i.e., the computation overhead of the whole system does not change)
or the server does not send out the offloading decision update request, the offloading decision is optimal.

\section{Simulation Results}

\begin{figure*}[h]
\centering\subfloat[Computation overhead of whole system.\label{fig:a}]{\includegraphics[width=8cm]{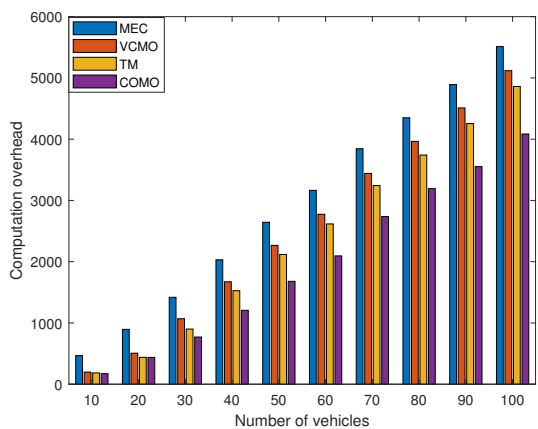}}\subfloat[Computation overhead of each vehicle.\label{fig:b}]{\includegraphics[width=8cm]{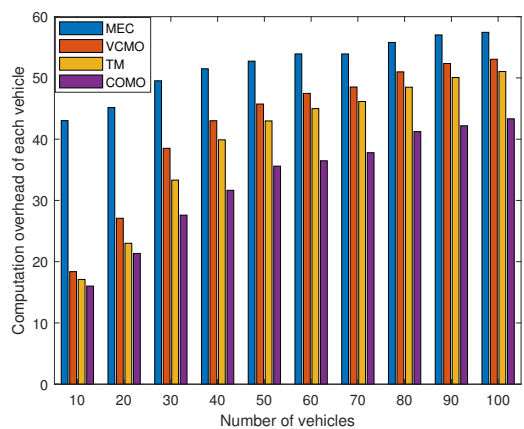}

}

\caption{Computation overhead under different numbers of vehicles. \label{fig:The2}}
\end{figure*}
\begin{figure*}[h]
\centering\subfloat[Average computation overhead.\label{fig:1}]{\centering\includegraphics[width=6cm]{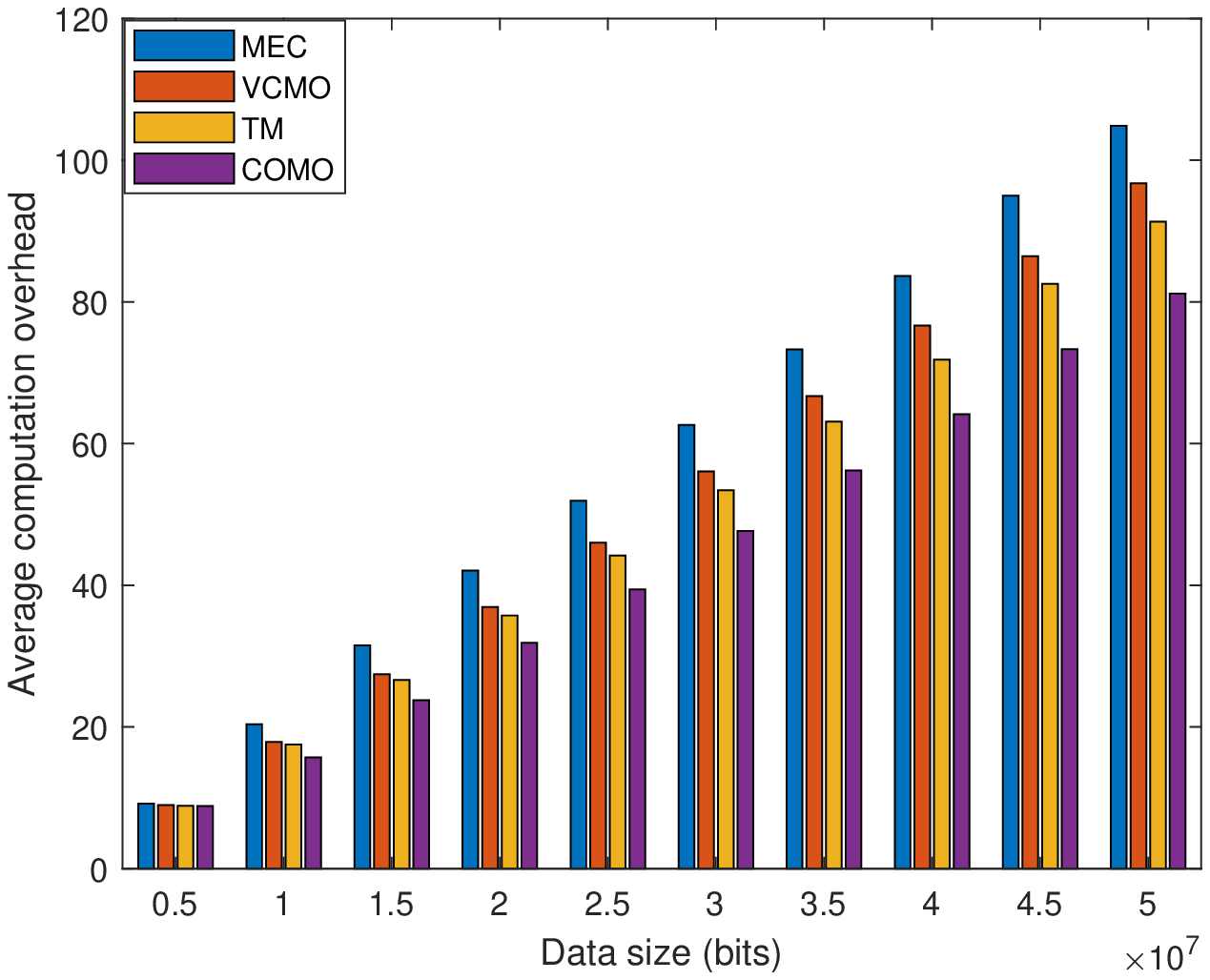}}\subfloat[Average delay.\label{fig:2}]{\centering\includegraphics[width=6cm]{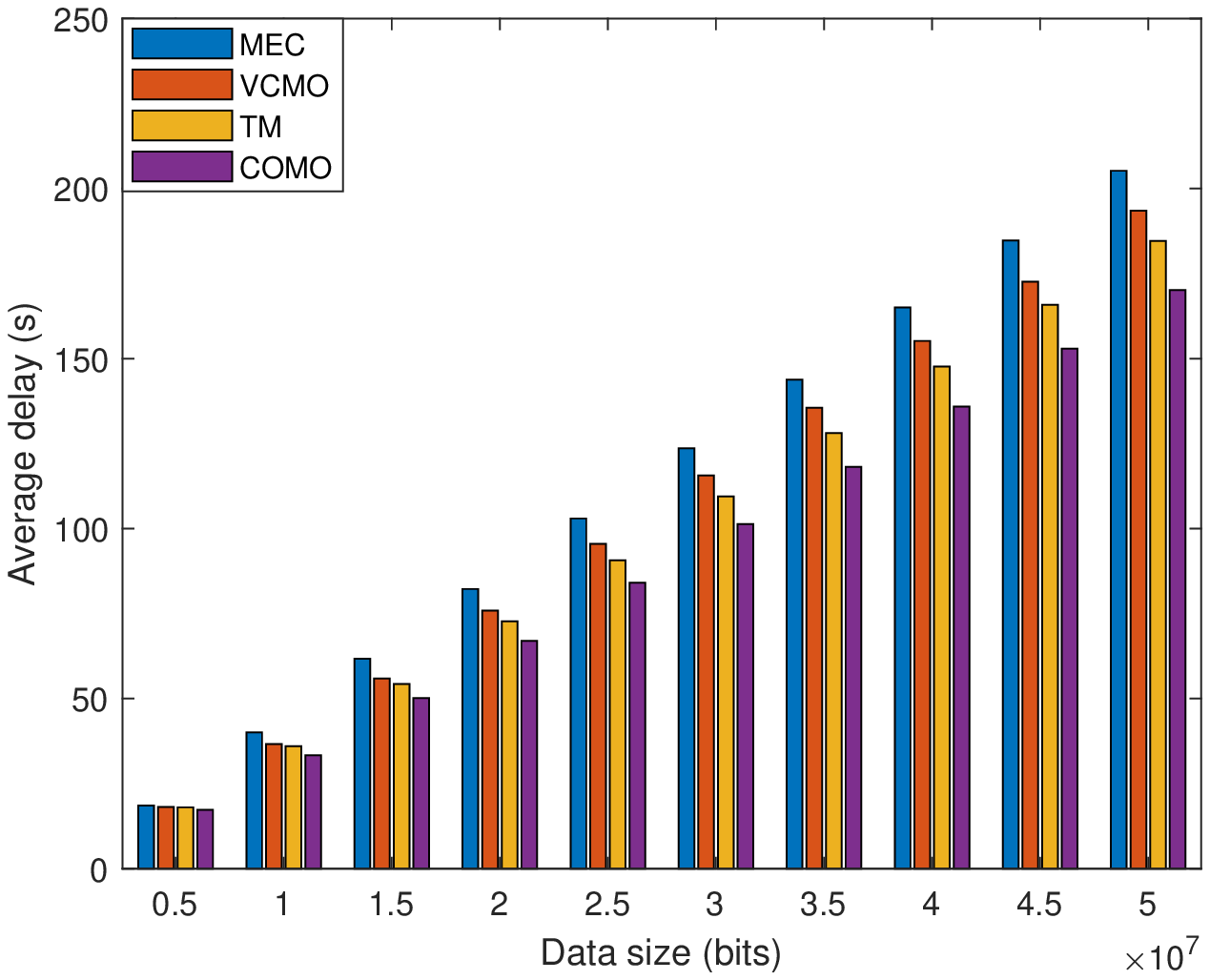}}\subfloat[Average energy consumption.\label{fig:3}]{

\centering\includegraphics[width=6cm]{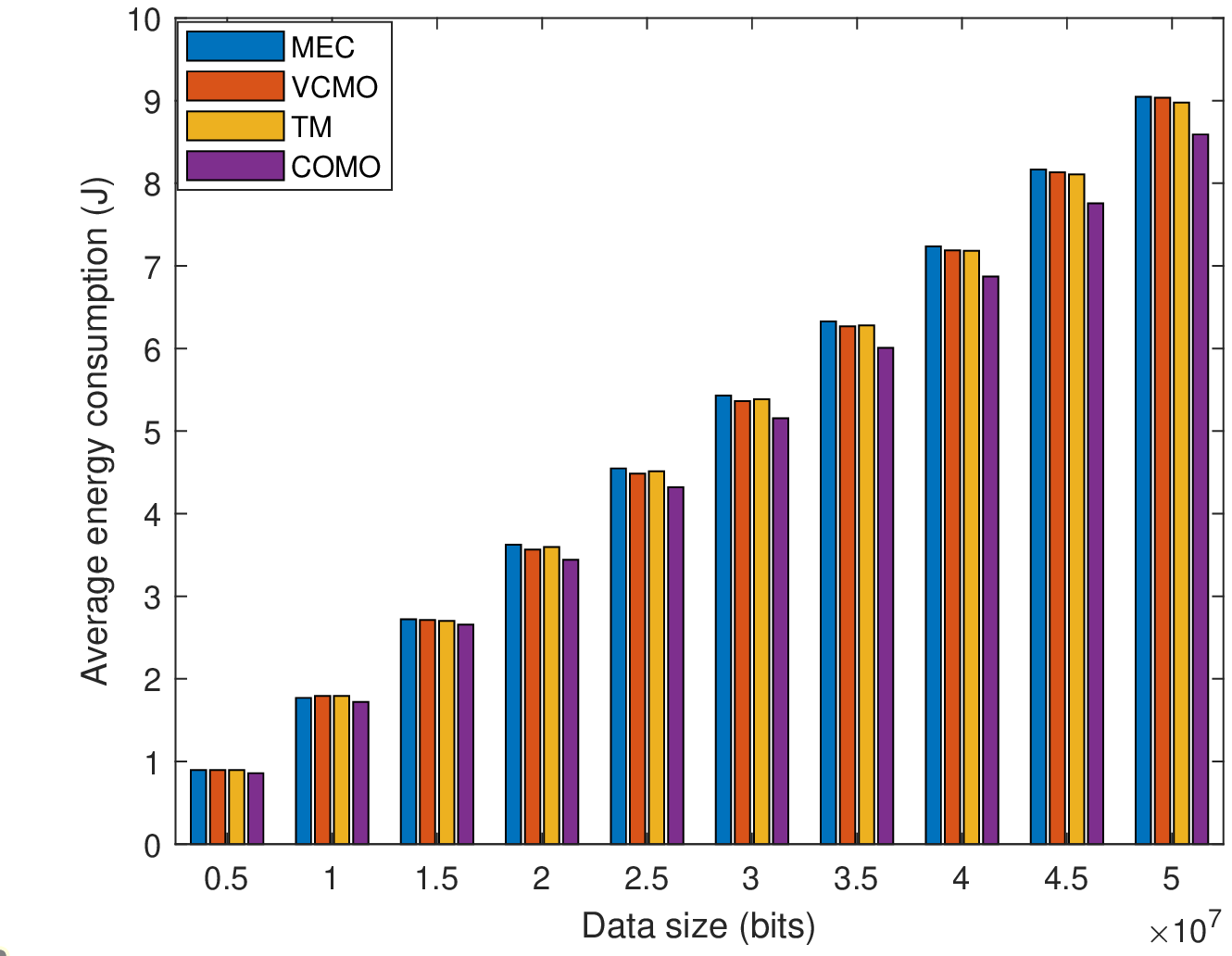}

}

\subfloat[Average computation overhead reduction rate.\label{fig:4}]{\centering\includegraphics[width=6cm]{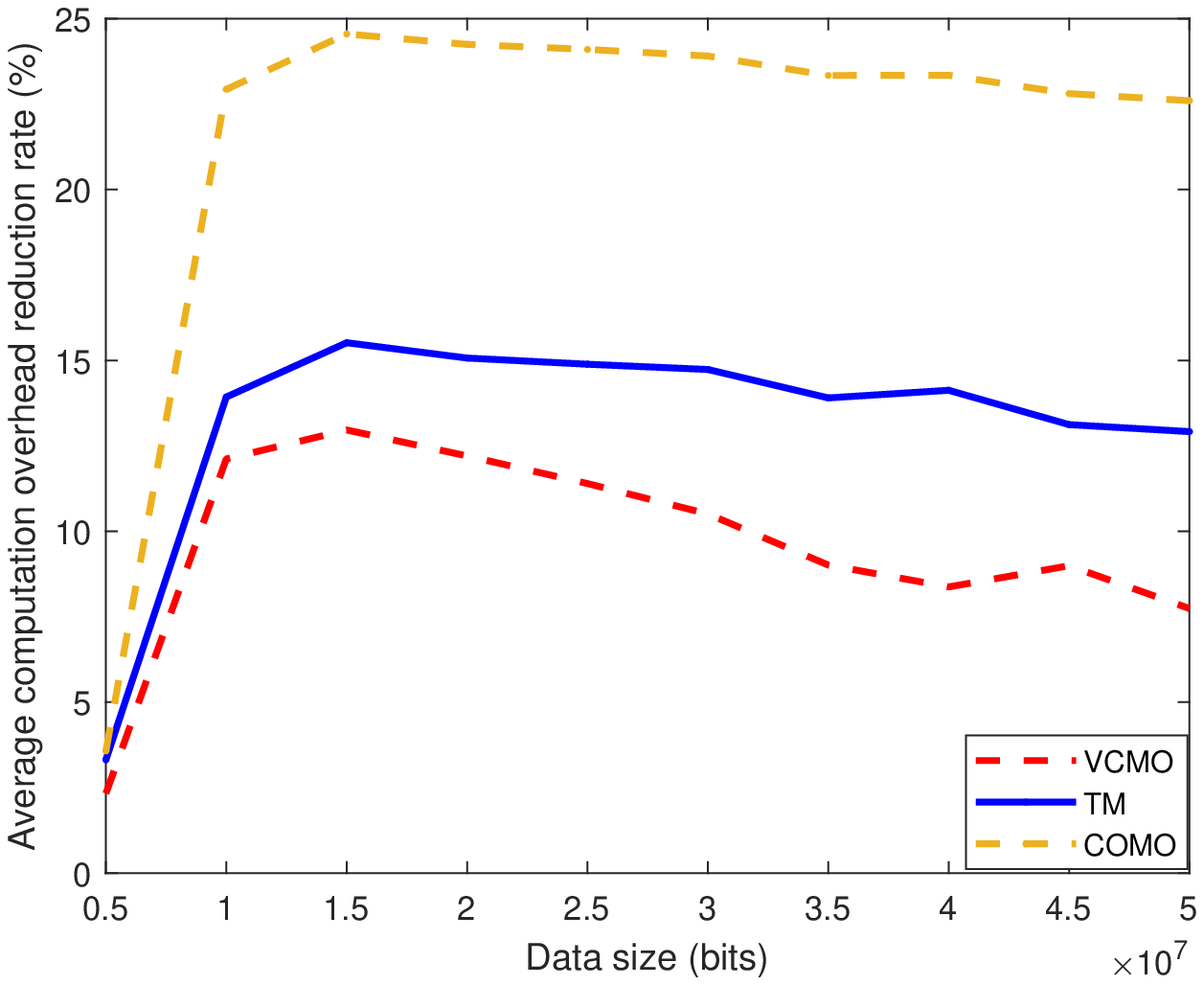}}\subfloat[Average delay reduction rate.\label{fig:5}]{\centering\includegraphics[width=6cm]{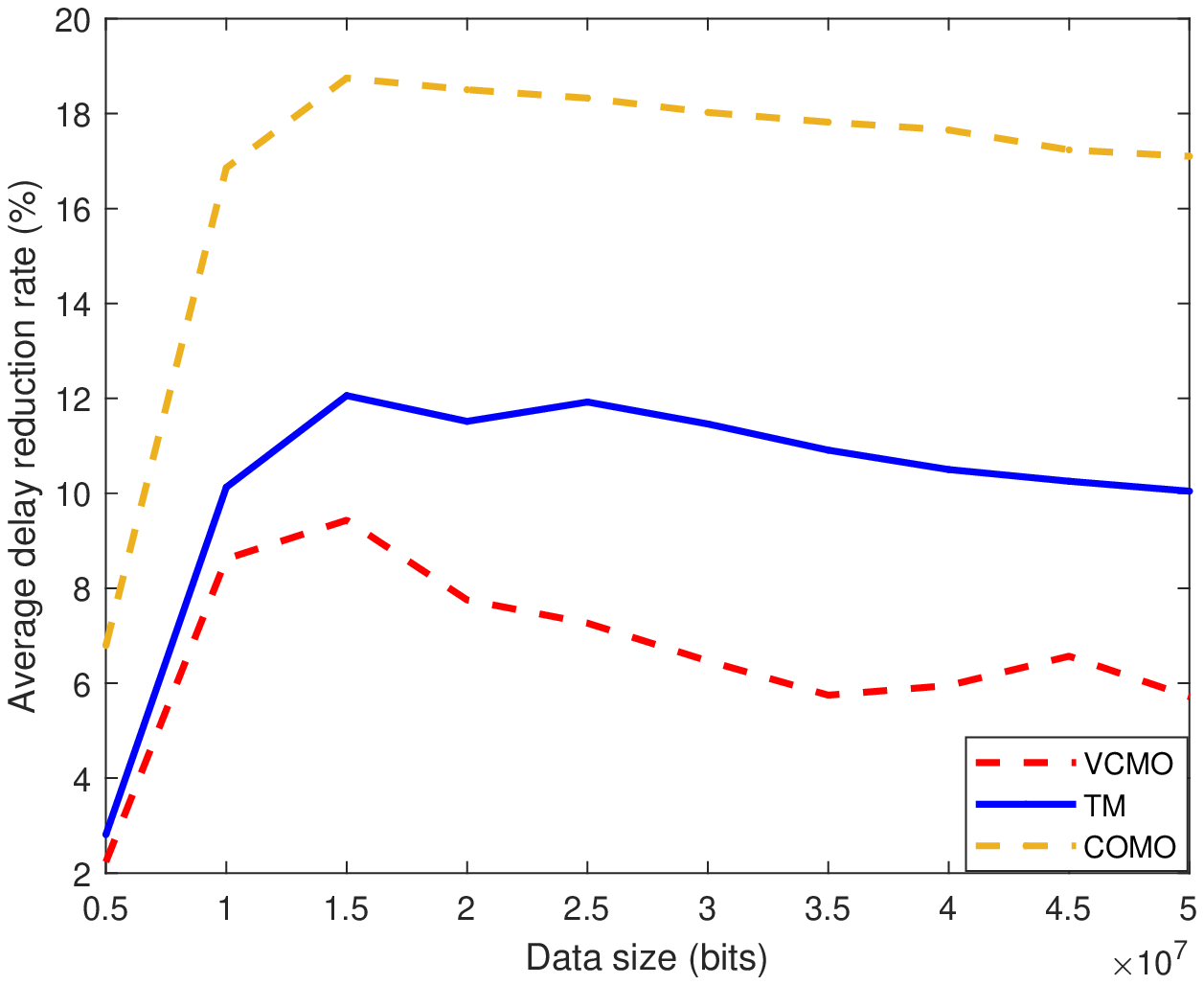}}\subfloat[Average energy consumption reduction rate.\label{fig:6}]{

\centering\includegraphics[width=6cm]{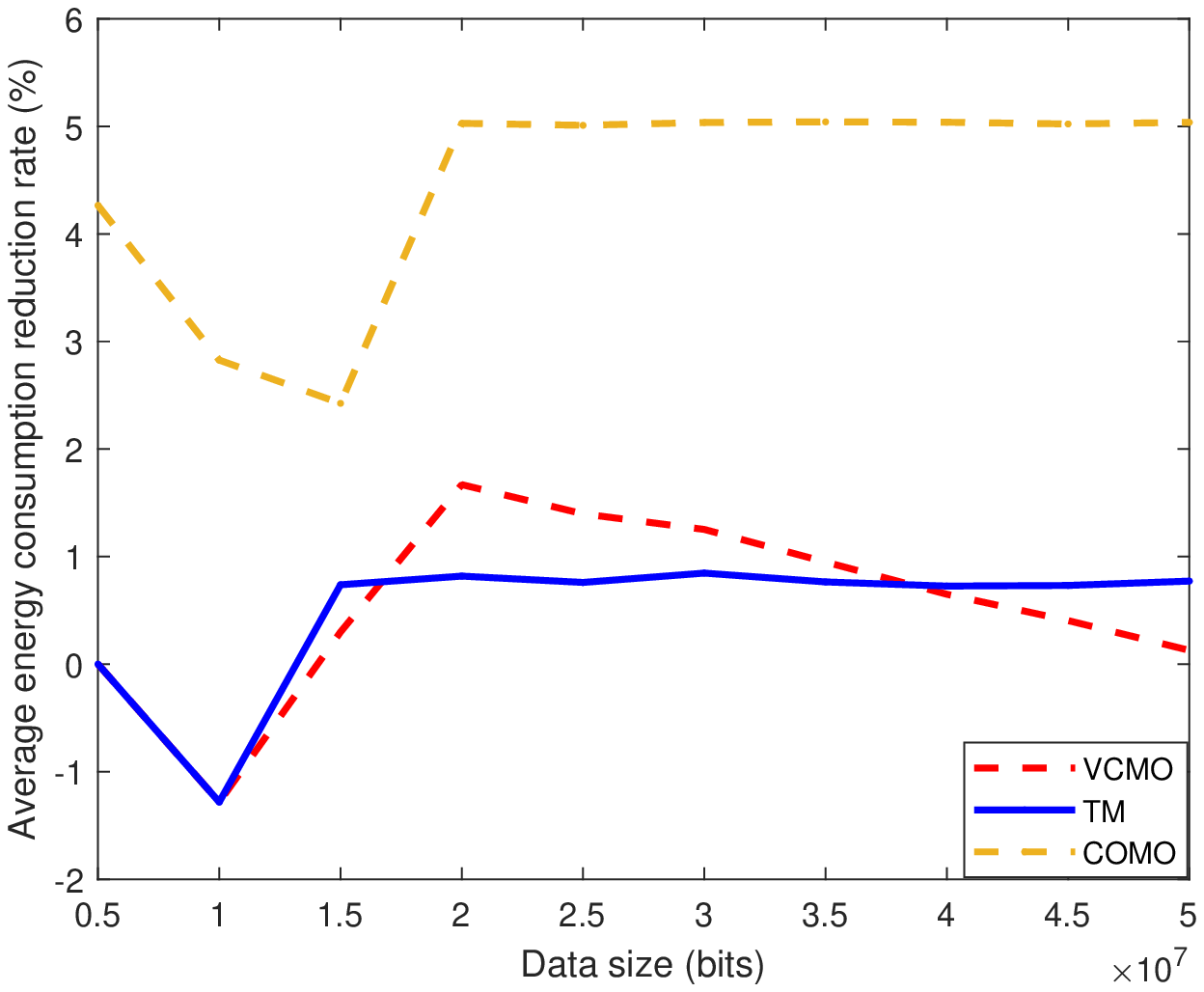}

}

\caption{Performance of different algorithms with different data sizes with
$M=100$.}
\end{figure*}

In this section, we provide simulation results to demonstrate the
performance of the proposed algorithms in the MEC-assisted vehicular
network system. RSUs are located on one-way roads. We use
urban mobility simulations to simulate road traffic. The speed of vehicles is $108\:\mathrm{km/h}$, and the maximum delay tolerance is
$\tau$ for the task. The simulation
parameters are shown in Table \ref{Table1}. Based on the most basic
no-migration task offloading scheme MEC to evaluate the efficiency
of the TM algorithm and the COMO algorithm, we also compare the proposed
algorithm with the VCMO algorithm \cite{37}.

Fig. \ref{fig:The2} shows the computation overhead of the whole system
and each vehicle under different numbers of vehicles. In this simulation,
we do not consider the maximum delay tolerance $\tau$.
Fig. \ref{fig:The2} shows that the computation
overhead of the whole system and that of each vehicle increases as the number
of vehicles increases. For the TM algorithm and the COMO algorithm, the computation overhead of the whole system and that of each vehicle are significantly
lower than those for the MEC and VCMO algorithms. In Fig. \label{fig:a}4(a), initially, the computation overhead of the VCMO algorithm, the TM
algorithm and the COMO algorithm are basically the same. As the number
of vehicles increases, the computation overhead gap between the COMO algorithm
and the VCMO algorithm becomes large. In Fig. \label{fig:a}4(b),
when the number of vehicles is small, the vehicle can offload the
task to a nearby RSU, and task migration is adopted
for the large amount of data; thus, the computation overhead of each vehicle
is low. As the number of vehicles increases, the co-channel interference
on offloading and migration links improves the overall computation
overhead, resulting in an increase in the computation overhead of each vehicle.

\begin{table*}
\caption{REDUCTION RATE \label{tab:REDUCTION-RATE}}

\centering%
\begin{tabular}{|c|c|c|c|}
\hline
\multicolumn{4}{|c|}{Average computation overhead reduction rate }\tabularnewline
\hline
\hline
& $5\mathrm{Mbits}$ & $15\mathrm{Mbits}$ & $50\mathrm{Mbits}$\tabularnewline
\hline
VCMO & $2.3\%$ & $13.0\%$ & $7.7\%$\tabularnewline
\hline
TM & $3.3\%$ & $15.5\%$ & $12.9\%$\tabularnewline
\hline
COMO & $3.6\%$ & $24.6\%$ & $22.6\%$\tabularnewline
\hline
\end{tabular}%
\begin{tabular}{|c|c|c|}
\hline
\multicolumn{3}{|c|}{Average delay reduction rate}\tabularnewline
\hline
\hline
$5\mathrm{Mbits}$ & $15\mathrm{Mbits}$ & $50\mathrm{Mbits}$\tabularnewline
\hline
$2.2\%$ & $9.4\%$ & $5.7\%$\tabularnewline
\hline
$2.8\%$ & $12.1\%$ & $10.0\%$\tabularnewline
\hline
$6.7\%$ & $18.7\%$ & $17.1\%$\tabularnewline
\hline
\end{tabular}%
\begin{tabular}{|c|c|c|c|c|}
\hline
\multicolumn{5}{|c|}{Average energy consumption reduction rate}\tabularnewline
\hline
\hline
$5\mathrm{Mbits}$ & $10\mathrm{Mbits}$ & $15\mathrm{Mbits}$ & $20\mathrm{Mbits}$ & $50\mathrm{Mbits}$\tabularnewline
\hline
$0.0\%$ & $-1.3\%$ & $0.3\%$ & $0.8\%$ & $0.8\%$\tabularnewline
\hline
$0.0\%$ & $-1.3\%$ & $0.7\%$ & $1.7\%$ & $0.1\%$\tabularnewline
\hline
$4.3\%$ & $2.8\%$ & $2.4\%$ & $5.0\%$ & $5.0\%$\tabularnewline
\hline
\end{tabular}
\end{table*}

Figs. \label{fig:1}5(a), \label{fig:2}5(b) and \label{fig:3}5(c) show the average
computation overhead, average delay and average energy consumption
versus the data size. We do not consider the
maximum delay tolerance $\tau$. From Figs. \label{fig:1}5(a), \label{fig:2}5(b) and \label{fig:3}5(c), we can clearly see that as the data size increases, the average computation overhead, the average delay
and the average energy consumption increases accordingly. Moreover, the COMO algorithm, TM algorithm and VCMO algorithm, which take into account task migration, perform better than MEC without considering task migration. When the data
size is large, the COMO algorithm performs best. Figs. \label{fig:4}5(d), \label{fig:5}5(e) and \label{fig:6}5(f) show the average computation overhead reduction rate, the average
delay reduction rate and the average energy consumption reduction
rate versus the data size. In the system, we define the reduction
rate as $(\Upsilon_{\mathrm{MEC}}(\boldsymbol{d})-\Upsilon_{\mathrm{VCMO/TM/COMO}}(\boldsymbol{d}))/\Upsilon_{\mathrm{MEC}}(\boldsymbol{d})$.
Some reduction rate data are shown in Table \ref{tab:REDUCTION-RATE}.
Initially, the size of the vehicle computing task data is small,
the number of vehicles applying task migration is small, and the reduction
rate of the average computation overhead is low. As the vehicle computing task data size increases, more vehicles need to
use task migration, and the reduction rate of the average computation
overhead increases rapidly from $3.6\%$ to $24.6\%$. When the vehicle computing task data size is larger than $15\mathrm{Mbits}$, the
reduction rate of the average computation overhead slowly decreases.
Accordingly, the average delay reduction rate increases from $6.7\%$
to $18.7\%$, and the average energy computation reduction rate increases
from $4.3\%$ to $5.0\%$. In Fig. \label{fig:6}5(f), the curve of the average energy consumption reduction rate shows an initial
decline in the rising trend. This is because if we consider only the energy consumption of vehicles, when the data size is in a certain range, the local energy consumption is slightly lower than other forms of energy consumption, but migration tasks cause the energy consumption of TM and VCMO
to be higher than that of MEC. As the data size increases,
local energy consumption exceeds other forms of energy consumption.
Then, the energy consumption of TM and VCMO is lower than that of
MEC.

\begin{figure}
\centering\includegraphics[width=8cm]{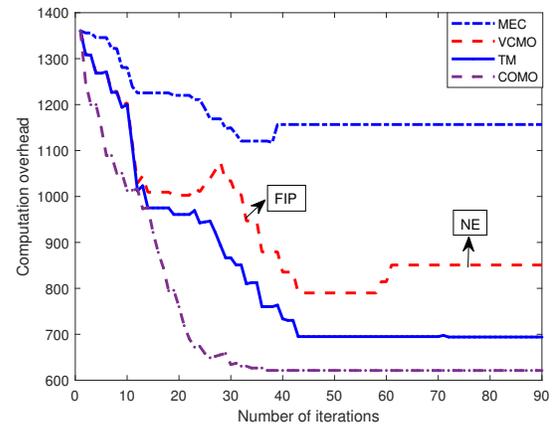}

\caption{Convergence behavior of different algorithms in terms of the system
overhead with $M=30$.\label{fig:Convergence-behavior-of}}
\end{figure}

Fig. \ref{fig:Convergence-behavior-of} shows the convergence behavior
of different algorithms when the system has 30 vehicles. In this simulation,
we do not consider the maximum delay tolerance $\tau$ for the task.
Based on the convergence and NE of the game, and
according to the FIP property, the game can go through a finite number of steps
to reach the NE. Clearly, all vehicles
choose to process tasks locally initially, so this is the point
where the computation overhead is highest. As the game proceeds, more tasks are offloaded to nearby vehicles and RSUs, and the computing
results are migrated via V2V or I2I migration links. Some bulges are observed on
the VCMO and TM curves because in the system, the vehicle is seen as the
player and every vehicle is aiming to minimize its own computation
overhead. Even if the vehicle minimizes its own computation overhead,
its offloading decision may result in an increase in the transfer
rate or migration rate, which in turn increases the overall computation
overhead, resulting in a bump in the curve. After a finite number
of iterations, the game reaches NE. Furthermore, as the game reaches NE, MEC has the highest computation overhead, followed by VCMO, TM and COMO.

\begin{figure}
\centering\includegraphics[width=8cm]{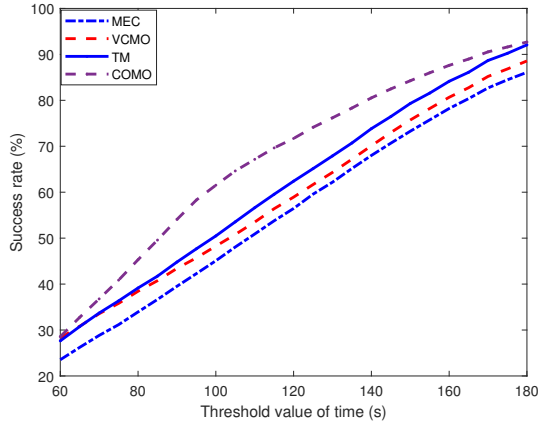}

\caption{Success rate of different algorithms with limited time threshold.
\label{fig:The-success-rate}}
\end{figure}

Fig. \ref{fig:The-success-rate} shows the success rate of different
algorithms with a time limit. The success rate refers to the percentage
of tasks completed within the time limit. We set the variable
$\mu_{m}$ randomly as $\left[0.5,1\right]$ and $\alpha_{m}$ is
randomly assigned from $\{0,0.5,1\}$ for vehicle $m$. The number of vehicles is
set to 100, that is, $M=100$. The success rate of different algorithms
increases almost linearly with an increase in the time limit. When the time limit is short, the success
rates of the VCMO algorithm, TM algorithm and COMO algorithm are very
low and tend to be the same due to the large size of the computing task.
Compared with the VCMO algorithm, within a certain range of time limits, the TM algorithm and COMO algorithm have high success rates. As the time limit increases, the success rates of different algorithms tend to be the same.  When the success rate is 100\%, the value of the maximum delay tolerance and the threshold value of time in Fig. \ref{fig:The-success-rate} are the same.

\section{Conclusion}

In this paper, we have studied an MEC-assisted vehicular network that considers task migration. In the system, we proposed
a TM algorithm to minimize vehicle computation overhead and a COMO
algorithm to minimize system computation overhead. We have also considered
the large amount of computing task data and the vehicle driving out
of the RSU communication range at fast speed. By means of a joint task migration
and task offloading model, we have modeled the offload decision problem
as a game and have demonstrated that the game always has an NE. Moreover, both algorithms can reach the NE in a finite
number of steps. Numerical simulations have shown that the proposed
TM and COMO algorithms can reduce the computation overhead and improve
the success rate.

\begin{appendices}
\section{Proof of Theorem 1}\label{prf1}
\begin{IEEEproof}
To reduce the computation overhead of the vehicle itself,
the vehicle selects the offloading scheme with the minimum computation
overhead from five offloading schemes. Consider two consecutive
iteration states $i$ and $i+1$, and assume that the offloading decision
$d_{m,i+1}$ was formed from $d_{m,i}$ after an iteration was performed.
Both operations occur if and only if the vehicle computation overhead
is decreased. It can be expressed as
\[
d_{m,i}\rightarrow d_{m,i+1}\Leftrightarrow\Omega_{m}(d_{m,i},\boldsymbol{d}_{-m})>\Omega_{m}(d_{m,i+1},\boldsymbol{d}_{-m}).
\]
Therefore, the computation overhead of vehicle $m$ is always decreasing,
as shown by
\[
d_{m,initial}\rightarrow d_{m,1}\rightarrow d_{m,2}\rightarrow\cdots\rightarrow d_{m,final},
\]
where $d_{m,initial}$ and $d_{m,final}$ are the initial and final
decisions of vehicle $m$, respectively. Therefore, the computation
overhead per vehicle is minimized. Since the number of vehicles is
finite and the offloading decisions of vehicles are also finite, the Nash
equilibrium can be guaranteed under a finite number of iterations.
\end{IEEEproof}
\end{appendices}

\begin{IEEEbiography}
[{\includegraphics[width=1in,height=1.25in,clip,keepaspectratio]{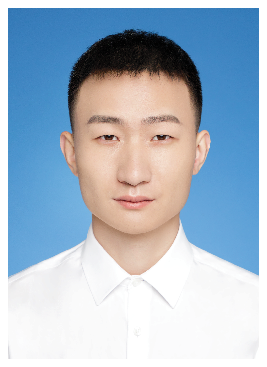}}]{Haipeng Wang}(Student Member, IEEE) received the B.S. degree from the China Jiliang University (CJLU), Hangzhou, China, in 2018. He is currently working toward the M.S. degree in communication engineering with the School of Information and Communication Engineering, Beijing University of Posts and Telecommunications (BUPT), Beijing, China. His current research interests include mobile edge computing (MEC), Internet of Vehicles (IoV) and task offloading.
\end{IEEEbiography}
\begin{IEEEbiography}
[{\includegraphics[width=1in,height=1.25in,clip,keepaspectratio]{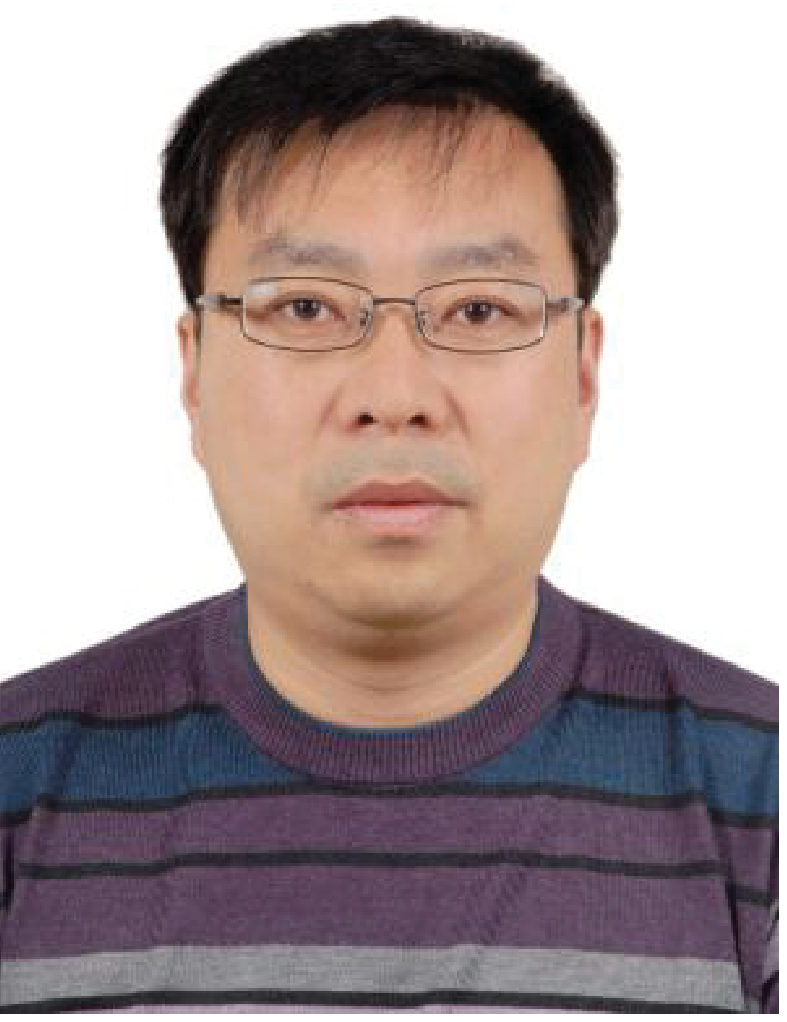}}]{Tiejun Lv}
(M'08-SM'12) received the M.S. and Ph.D. degrees in electronic engineering from the University of Electronic Science and Technology of China (UESTC), Chengdu, China, in 1997 and 2000, respectively. From January 2001 to January 2003, he was a Postdoctoral Fellow with Tsinghua University, Beijing, China. In 2005, he was promoted to a Full Professor with the School of Information and Communication Engineering, Beijing University of Posts and Telecommunications (BUPT). From September 2008 to March 2009, he was a Visiting Professor with the Department of Electrical Engineering, Stanford University, Stanford, CA, USA. He is the author of three books, more than 100 published IEEE journal papers and 200 conference papers on the physical layer of wireless mobile communications. His current research interests include signal processing, communications theory and networking. He was the recipient of the Program for New Century Excellent Talents in University Award from the Ministry of Education, China, in 2006. He received the Nature Science Award in the Ministry of Education of China for the hierarchical cooperative communication theory and technologies in 2015.
\end{IEEEbiography}
\begin{IEEEbiography}
[{\includegraphics[width=1in,height=1.25in,clip,keepaspectratio]{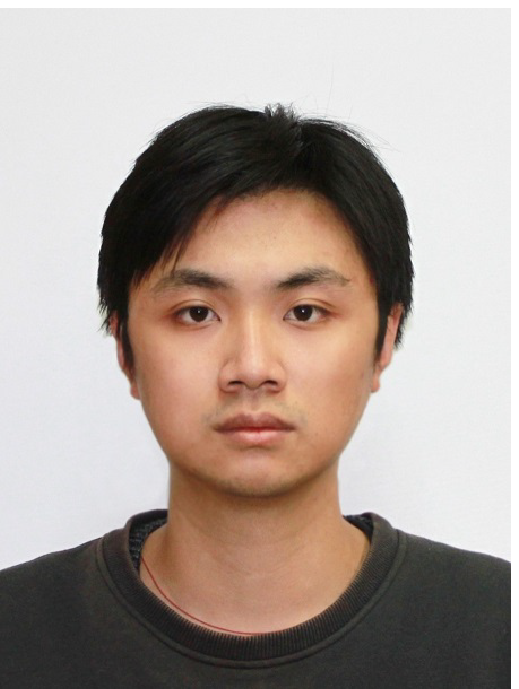}}]{Zhipeng Lin}
(Member, IEEE) received the Ph.D. degrees from the School of Information and Communication Engineering, Beijing University of Posts and Telecommunications, Beijing, China, and the School of Electrical and Data Engineering, University of Technology of Sydney, NSW, Australia, in 2021. Currently, He is an Associate Researcher in the College of Electronic and Information Engineering, Nanjing University of Aeronautics and Astronautics, Nanjing, China. His current research interests include signal processing, massive MIMO, spectrum sensing, and UAV communications.
\end{IEEEbiography}
\begin{IEEEbiography}
[{\includegraphics[width=1in,height=1.25in,clip,keepaspectratio]{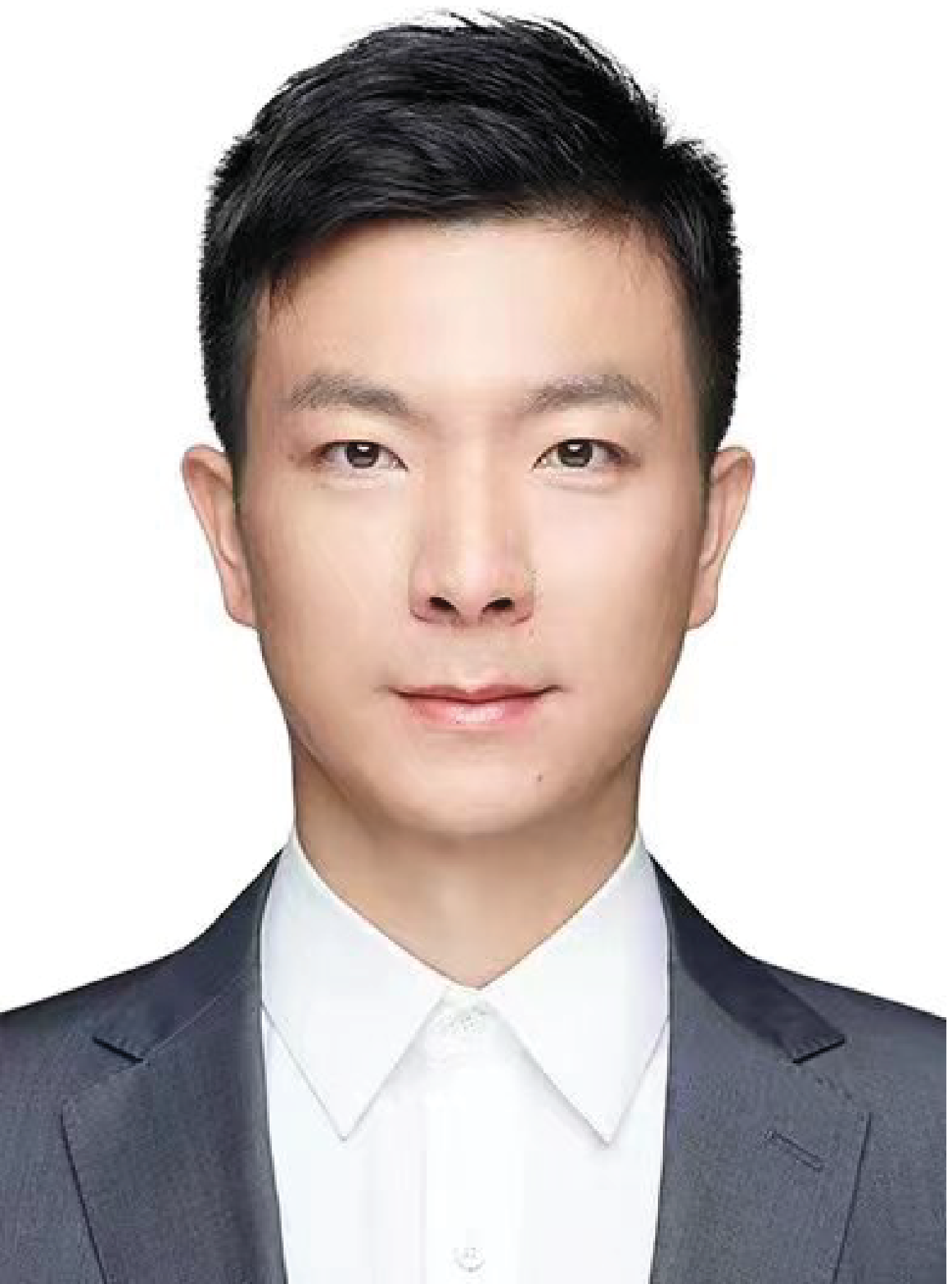}}]{Jie Zeng} (M'09--SM'16) received the B.S. and M.S. degrees from Tsinghua University in 2006 and 2009, respectively, and received two Ph.D. degrees from Beijing University of Posts and Telecommunications in 2019 and the University of Technology Sydney in 2021, respectively.

From July 2009 to May 2020, he was with the Research Institute of Information Technology, Tsinghua University. From May 2020 to April 2022, he was a postdoctoral researcher with the Department of Electronic Engineering, Tsinghua University. Since May 2022, he has been an associate professor with the School of Cyberspace Science and Technology, Beijing Institute of Technology.

His research interests include 5G/6G, URLLC, satellite Internet, and novel network architecture. He has published over 100 journal and conference papers, and holds more than 40 Chinese and international patents. He participated in drafting one national standard and one communication industry standard in China.

He received Beijing's science and technology award of in 2015, the best cooperation award of Samsung Electronics in 2016, and Dolby Australia’s best scientific paper award in 2020.
\end{IEEEbiography}
\end{document}